\documentclass[twocolumn, 10pt, english,
superscriptaddress, aps, prl,amsmath,amssymb,floatfix,longbibliography
]{revtex4-2}
\usepackage[margin=1.5cm]{geometry}
\usepackage[T1]{fontenc}
\usepackage[utf8]{inputenc}
\usepackage{color}
\usepackage[english]{babel}
\usepackage{amsmath}
\usepackage{graphicx}
\usepackage{cancel}
\usepackage{esint}
\usepackage{float}
\usepackage{tikz}
\usepackage{empheq}
\usepackage{aligned-overset}
\usepackage{amsmath,amssymb}
\usepackage{microtype}
 \usepackage{subfiles}
\usepackage{booktabs}
\usepackage{mathrsfs}

\usepackage[export]{adjustbox}
\usepackage[unicode=true,pdfusetitle,bookmarks=true,bookmarksnumbered=false,bookmarksopen=false,breaklinks=false,pdfborder={0 0 0},pdfborderstyle={},backref=false,colorlinks=true]{hyperref}

\makeatletter

\usepackage{dcolumn}
\usepackage{bm}
\usepackage{color}
\usepackage{soul}
\usepackage{amsfonts}
\usepackage{slashed}
\usepackage{enumerate}
\usepackage{mathtools}
\usepackage{physics}
\usepackage{amsthm}
\usepackage{inputenc}
\usepackage[normalem]{ulem}
\usepackage{svg}
\usepackage{comment}

\let\old@makecaption=\@makecaption
\usepackage{subcaption}
\usepackage{graphicx}
\let\@makecaption=\old@makecaption

\usepackage[figurename=FIG.]{caption}

\usepackage{microtype}

\theoremstyle{remark}

\newcommand{\diag}{\operatorname{diag}}

\usepackage{epsfig}
\usepackage{dsfont}
\usepackage{xcolor}

\makeatother
\begin{document}

\title{What a chaotic Floquet system remembers:\\ Universal Krylov spread and an endpoint transition}

\author{E. Medina-Guerra}
\email{edward.medinaguerra@weizmann.ac.il}
\affiliation{Department of Condensed Matter Physics, Weizmann Institute of Science, Rehovot 7610001, Israel}
\author{I. V. Gornyi}
\affiliation{\mbox{Institute for Quantum Materials and Technologies, Karlsruhe Institute of Technology, 76131 Karlsruhe, Germany}}
\affiliation{\mbox{Institut für Theorie der Kondensierten Materie, Karlsruhe Institute of Technology, 76131 Karlsruhe, Germany}}

\author{Yuval Gefen}
\affiliation{Department of Condensed Matter Physics, Weizmann Institute of Science, Rehovot 7610001, Israel}

\date{\today}

\begin{abstract}
A chaotic Floquet system can remember its initial state by favoring early Krylov directions at late times. 
For circular ensembles of Dyson index $\beta$ and symmetry-matched random initial states, the stationary mean 
Krylov position (spread complexity) grows as $P(\beta)D$, with Krylov dimension $D$ and closed-form $P(\beta)<1/2$. 
As $D\to\infty$, the mean last-site occupation in units of $1/D$ tends to zero for $\beta\leq2$ and to $1-2/\beta$ 
for $\beta>2$. One-dimensional localization explains this transition. Kicked tops and Floquet circuits closely 
follow the finite-size predictions.
\end{abstract}
\maketitle

\paragraph{Introduction.}
Consider a wave packet released at one site of a disordered sample. 
Its long-time probability at a distant site depends on whether the same 
eigenfunctions have appreciable intensity at both the distant site and 
the source. Anderson localization suppresses this probability at distances 
beyond the localization length~\cite{EversMirlin2008}. 
In a periodically driven quantum system, the same question can be asked 
without choosing positions in real space. Let $U$ be the evolution operator 
over one period of the drive (the Floquet operator) and $\ket{\psi_0}$ 
the initial state. Sampling the state after each period 
gives the ordered sequence $\ket{\psi_t}=U^t\ket{\psi_0}$, $t=0,1,2,\ldots$, called 
the \emph{Floquet orbit} of the initial state. 
The component of $U\ket{\psi_0}$ orthogonal to $\ket{\psi_0}$ defines 
a new direction in Hilbert space. The component of $U^2\ket{\psi_0}$ 
orthogonal to both preceding directions defines the next one. 
Continuing in this way gives the Krylov basis 
$\ket{K_0}=\ket{\psi_0},\ket{K_1},\ldots,\ket{K_{D-1}}$, 
ordered by the first period at which each direction 
becomes accessible~\cite{bala,PhysRevE.108.054222,PhysRevE.110.034201,PhysRevE.111.014220,Nandy_2025}. 
We picture these directions as the sites of a chain. The initial state sits at site zero, 
and the evolving state spreads along the chain as a wave packet spreads through a sample. 
When a chaotic quantum system equilibrates locally, what memory of its
preparation survives in the way it explores Hilbert space?

The same chaotic drive that spreads the state along the Krylov chain can also heat an 
isolated many-body system toward infinite temperature. Thermalization is usually tested 
with local observables defined independently of the initial state~\cite{I1,I2,I3,I4,I10}. 
The eigenstate thermalization hypothesis says that, within a symmetry sector, the expectation 
values of these observables in typical Floquet eigenstates 
are close to those at infinite temperature~\cite{PolkovnikovEtAl2011,I5,I12,I13}. 
For initial states that populate these eigenstates, the long-time averages are nearly 
independent of the preparation. Prethermalization at high drive frequency 
and many-body localization delay or prevent this loss of 
local memory~\cite{BukovDAlessioPolkovnikov2015,I15,I16,I14,I17,I18,I19}.
Krylov methods
in operator space have also been used to study the relaxation of local Floquet
observables~\cite{3vgd-y39w}.

In this Letter, quasienergies are expressed in phase units as the eigenphases of $U$. Within a symmetry 
sector, spectral chaos is characterized by Wigner--Dyson correlations of these eigenphases, whose universal 
form is fixed by the symmetry class~\cite{dyson1962statistical,I7,I8,haake2018quantum,I22,I23,I24,I25,I26}. 
Throughout, ``chaotic'' refers to these spectral correlations; it does not imply a uniform occupation 
of the Krylov basis. Stroboscopic evolution preserves the weight of the state in each Floquet eigenspace. 
The Krylov basis is itself built from the drive and the initial state, so that its stationary occupations 
depend jointly on these weights and on the quasienergies, which makes them hard to predict. In Hamiltonian 
dynamics, the role of the weights in setting the late-time spread has been studied in Refs.~\cite{competitor1,p1}. 
A scaling argument supported by numerical results was put forward in Ref.~\cite{streamed}, which 
predicted that the late-time Krylov spread in chaotic Floquet systems is proportional to the chain length. 
Neither this proportionality nor its coefficient has been derived. Moreover, an extensive spread alone does 
not tell whether the most distant directions receive their share of the probability.

We model chaotic quasienergies by Dyson's circular ensembles, in which the index $\beta$ enters at three 
levels. First, it sets the strength of the level repulsion~\cite{dyson1962statistical,haake2018quantum}, 
and symmetry allows only three values in a chaotic drive: $\beta = 1,2,4$ for the orthogonal, unitary, 
and symplectic classes. In the orthogonal and symplectic classes, time reversal squares to $+1$ and $-1$, 
respectively; breaking time reversal gives the unitary class. We pair each phase law with a matching 
distribution of initial weights, proportional to squares of real amplitudes for $\beta = 1$, squared 
moduli of complex amplitudes for $\beta = 2$, and sums of two squared complex moduli for $\beta = 4$ 
(End Matter Sec.~\ref{em:ensemble}); thus, $D$ counts Kramers pairs in the symplectic class. 
Second, the joint model of phases and weights extends to all real $\beta > 0$, which allows us to follow the 
results continuously through $\beta = 2$, the value of the unitary class. Third, this extension is implemented 
via a chain of scatterers described below, whose reflection strength is set to $\eta^2 = 2/\beta$ 
(a flux-driven ring realizes the same reflection law through Landau-Zener
transitions~\cite{GefenThouless1987,BlatterBrowne1988}).

\begin{figure*}[t!]
 \centering
 \includegraphics[width=\textwidth]{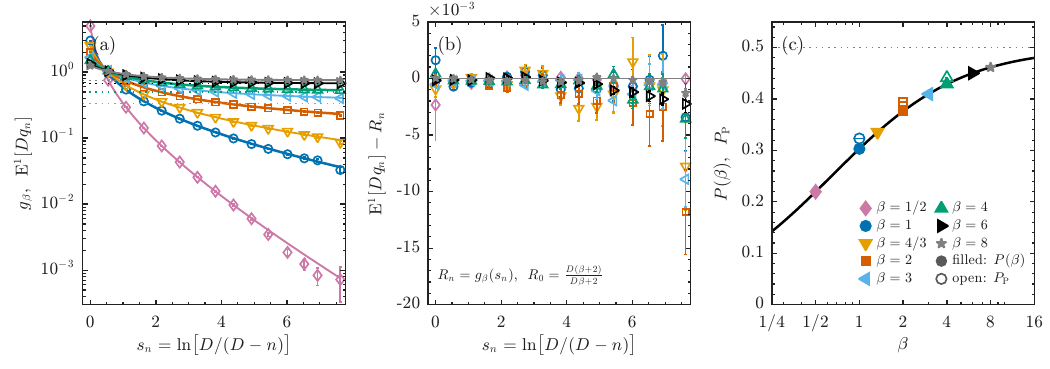}
 \caption{Stationary occupation of the Krylov chain and its mean position. 
 (a) Stationary occupation at site $n$, averaged over drives and divided by the uniform share $1/D$. 
 The symbols give $D\,\mathrm E[\bar p_n]$ at $D=2048$ and the curves give the limiting profile $g_\beta$. 
 The axis label $\mathrm E^1[Dq_n]$ expresses the same average by sampling one eigenmode with its initial 
 weight (End Matter Sec.~\ref{em:palm}). The logarithmic coordinate $s_n=\ln[D/(D-n)]$ resolves sites near 
 the far end. The eight values of $\beta$ are identified in (c). 
 Dotted lines give the endpoint limit $1-2/\beta$ for $\beta>2$. 
 (b) Data minus $R_n$, the limiting density with the exact finite-$D$ value used at the initial site. 
 Error bars give one standard error over 200 matrices; residuals at different sites share the same 
 matrices and are correlated. 
 (c) Scaled mean position $P(\beta)$ from Eq.~\eqref{eq:F} (curve with filled symbols at the eight values of $\beta$)
 and its counterpart $P_{\mathrm P}$ for independent uniform phases with the same overlap statistics (open symbols).}
 \label{fig:A1.0}
\end{figure*}

For these ensembles with symmetry-matched random initial states, we derive the large-$D$ limit of 
the ensemble-averaged stationary Krylov profile. First, the mean position along the chain, or spread complexity, 
grows as $P(\beta)D$. The coefficient $P(\beta)<1/2$ has a closed form and is universal: it depends only 
on the symmetry class. Second, the last site shows a transition that the mean position, smooth through $\beta=2$, 
does not. Specifically, as the chain grows, the last site retains a nonzero fraction of
its uniform share $1/D$ only for $\beta>2$.
One-dimensional localization on the Krylov chain explains this threshold, which corresponds to $\eta^2= 1$.

\paragraph{A wave packet on the Krylov chain.}
Let $p_n(t)=|\langle K_n|U^t|\psi_0\rangle|^2$ be the probability at site $n$ after $t$ periods, and 
let $\bar p_n$ be its average over infinitely many periods. Write $\varphi_a(n)=\langle K_n|z_a\rangle$ 
for the wave function on the chain of the Floquet eigenmode $\ket{z_a}$, whose eigenvalue is $z_a=e^{i\theta_a}$. 
Its weight in the initial state is $w_a=|\varphi_a(0)|^2$. We work within a single symmetry sector; 
$D$ denotes the number of distinct quasienergies with nonzero weight. 
For a degenerate quasienergy, $\ket{z_a}$ is the normalized projection of the initial state onto that eigenspace. 
Time averaging removes the interference between different quasienergies and gives
\begin{equation}\label{eq:cesaro}
 \bar p_n=\sum_{a=1}^D \abs{\varphi_a(0)}^2\abs{\varphi_a(n)}^2 .
\end{equation}

Each eigenmode contributes to the long-time probability at site $n$ in proportion to the product of 
its intensity at the starting site and its intensity at $n$. The same product governs the long-time 
spread of a wave packet over a disordered lattice~\cite{EversMirlin2008}. 
To read Eq.~\eqref{eq:cesaro} for a given drive, select an eigenmode with probability $w_a$. 
Averaging the evolving density matrix over time cancels the oscillating
terms between distinct quasienergies and gives
$\rho_{\mathrm d}=\sum_a w_a\ket{z_a}\bra{z_a}$.
The occupations $\bar p_n$ are its diagonal elements in the Krylov basis.
Its eigenvalues on the Krylov subspace are the weights $w_a$, so that the
long-time return probability $\bar p_0=\sum_a w_a^2$ is also the purity
$\operatorname{Tr}(\rho_{\mathrm d}^2)$.
This sum takes its minimum $1/D$ precisely for equal weights $w_a=1/D$.
In that case, $\rho_{\mathrm d}$ is the identity divided by $D$ on the
Krylov subspace. It is therefore maximally mixed there, and every Krylov
direction has occupation $1/D$ for any quasienergies.
Away from the starting site, the stationary occupations for fluctuating weights also depend on 
the quasienergy correlations~\cite{competitor1,suppression}.

\paragraph{Mean position and endpoint transition.} 
Figure~\ref{fig:A1.0}(a) shows the mean occupation in units of the uniform 
share $1/D$. 
As $D\to\infty$ at fixed $x=n/D$, it tends to a limiting density $g_\beta(x)$, which we obtain 
explicitly as an integral~\eqref{eq:density} in End Matter (Sec.~\ref{em:moments}). 
The density starts at $g_\beta(0)=1+2/\beta$, the mean return probability in units of $1/D$,
and decreases toward the far end. For a single realization, the center of mass of the discrete
occupation is the late-time Krylov spread complexity (the plateau of the spread), 
$\bar C_D=\sum_n n\bar p_n$~\cite{bala,Krylov_distribution,Krylov_distribution_2}, 
the standard measure of how far a state has spread along the Krylov chain. 
We find
\begin{equation}\label{eq:universal}
 \lim_{D\to\infty}\frac{\mathrm E[\bar C_D]}{D} =P(\beta)=\int_0^1 x\,g_\beta(x)\,\dd x ,
\end{equation}
with the closed form (Supplemental Material)
\begin{align}
 P(\beta)&=1-F\Bigl(1+\frac{\beta}{2},\beta\Bigr),\label{eq:F}\\
 F(\mu,q)&=\frac{\mu}{b(b+1)}\, {}_3F_2\!\left(\begin{matrix}1,b,\mu+1\\b+1,b+2\end{matrix};1\right),\nonumber
\end{align}
where $b=[\mu+(\mu^2+2q)^{1/2}]/2$ and ${}_3F_2$ is the generalized hypergeometric function.
For the orthogonal, unitary, and symplectic classes, respectively, Eq.~\eqref{eq:F} yields
\begin{equation}\label{eq:class-plateaus} 
P(1)\approx 0.3031,\ \ P(2)\approx 0.3762,\ \ P(4)\approx 0.4289.
\end{equation}
For a uniformly occupied chain, the mean position would be $(D-1)/2$. 
The circular ensembles give a smaller mean position for every $D>1$ (End Matter
Sec.~\ref{em:palm}). Figure~\ref{fig:A1.0}(c) shows how $P(\beta)$ approaches $1/2$ 
as level repulsion grows. Finite-chain calculations support the correction 
$\mathrm E[\bar C_D]=P(\beta)D-1/2+o(1)$ used in the comparisons below.

\begin{figure}[t!]
 \centering
 \includegraphics[width=\linewidth]{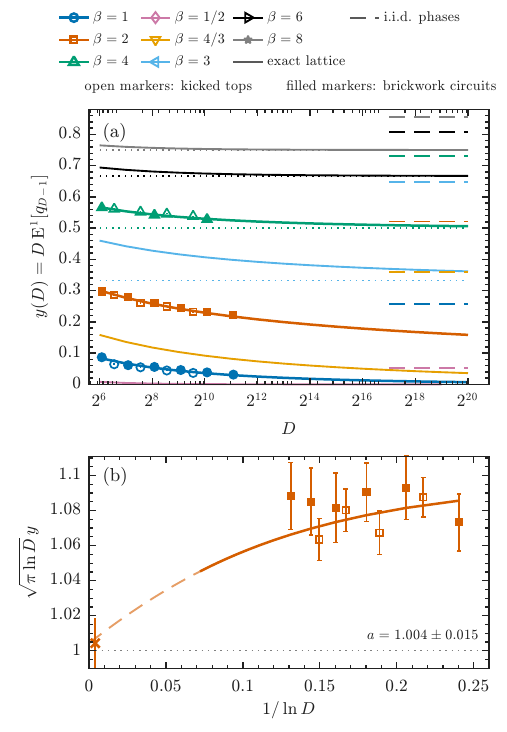}
 \caption{Occupation $y(D)=D\,\mathrm E[\bar p_{D-1}]$ of the last Krylov site in units of $1/D$. 
 (a) The exact random-matrix chain (curves), kicked tops (open symbols), and circuits (filled symbols).
 Dotted lines give the limit $1-2/\beta$ for $\beta>2$. 
 Dashed segments show independent phases with the same weight statistics. 
 Model error bars are one standard error over operators. 
 (b) At $\beta=2$, multiplying $y(D)$ by $\sqrt{\pi\ln D}$ tests the continuum prediction of
logarithmically slow decay. The proposed asymptote gives one on this scale. The solid curve 
is a fit $a+b/\ln D+c/\ln^2D$ to sizes $D\geq2^{14}$. Its extrapolation is dashed, and its 
intercept $a$ is the cross, with a bar giving the spread over fit variants.}
 \label{fig:A2}
\end{figure}

The mean position varies smoothly through $\beta=2$, but the last site of the chain
distinguishes two cases. With its occupation 
in units of $1/D$ $y(D)=D\,\mathrm E[\bar p_{D-1}]$, we prove
\begin{equation}\label{eq:end}
 \lim_{D\to\infty} y(D)=
 \begin{cases}
  0, & 0<\beta\leq 2,\\[2pt]
  1-2/\beta, & \beta>2 .
 \end{cases}
\end{equation}
For the matched ensembles,
$y(\infty) \equiv \lim_{D\to \infty}y(D) = 1/\lim_{D\to \infty}\mathrm{E}[(Dw_a)^{-1}]$.
The overlap density scales as $w^{\beta/2-1}$ near zero, so that the inverse
moment is finite only for $\beta>2$. The scaled occupation $D\bar p_{D-1}$
converges in probability to $y(\infty)$ over draws of the drive and
initial state.
For the orthogonal class, the end is progressively depleted as $D$ grows. 
For the unitary class, the continuum theory suggests a logarithmically slow depletion,
$y(D)\simeq(\pi\ln D)^{-1/2}$. Even at $D=2^{20}$, the last site retains about $16\%$ of 
a uniform share [Fig.~\ref{fig:A2}(a)]. For the symplectic class, the last direction 
retains half a uniform share on average, $\mathrm E[\bar p_{D-1}]\sim1/(2D)$. 
Every site has a positive occupation at finite $D$. The transition concerns the occupation 
of the site relative to $1/D$ at the remote end of an extensively occupied chain.

\paragraph{Why localization physics enters.} We can represent the drive by a chain 
of lossless beam splitters~\cite{Simon2005Orthogonal,CMV}. Each splitter mixes 
two neighboring amplitudes. In this representation, every Floquet eigenmode has 
the same intensity at each site as in the Krylov basis (Supplemental 
Sec.~\ref{sm:arnoldi-cmv}). For the circular ensembles, different links have 
independent reflection amplitudes~\cite{Killip}. At a distance $r\gg1$ from the far end, 
the mean reflection probability is asymptotic to $\eta^2/r$, 
with $\eta^2=2/\beta$~\cite{Killip,Killip1,streamed}.

Select an eigenmode with its initial weight. Across each link, its unnormalized intensity
acquires an independent random factor. Products of such factors build up large 
fluctuations, as in one-dimensional localization~\cite{EversMirlin2008,KiselevLastSimon1998}.
The $1/r$ law gives the same increase in the variance of the log intensity each 
time the remaining distance to the far end is halved. Hence, the natural coordinate 
is $s=-\ln(1-x)$ with $x=n/D$, as in Fig.~\ref{fig:A1.0}(a).
On this scale, the logarithm of the intensity performs Brownian motion with 
a constant downward drift. If $B(s)$ is a standard Brownian motion, the
intensity of the sampled mode per unit $x$ has the limiting profile
\begin{equation}\label{eq:gibbs}
 \mathcal G_\beta(x)=\frac{1}{\mathcal Z_\infty}
 \exp\!\left[\frac{2}{\sqrt{\beta}}B(s)-\frac{2}{\beta}s\right],
 \qquad s=-\ln(1-x).
\end{equation}
Here, $s$ measures position along the chain. Each Brownian intensity 
profile is normalized to unit total probability before averaging to 
obtain $g_\beta(x)$. Related Brownian profiles describe eigenfunction
intensities in one-dimensional systems at critical
disorder~\cite{Rifkind_2018,nakano2023shape}.

For a typical eigenmode, the downward drift in Eq.~\eqref{eq:gibbs} wins over 
Brownian fluctuations and the intensity fades toward the end. The mean is sensitive 
to rare eigenmodes with large intensity there~\cite{EversMirlin2008,I11}. Weighting 
the Brownian paths by their unnormalized intensity at a distant site reverses the drift 
up to that site (End Matter Sec.~\ref{em:endpoint}). On the reweighted paths, 
the typical intensity grows as $e^{\eta^2s}$, while the number of sites per unit $s$ 
shrinks as $e^{-s}$. For $\eta^2>1$, the total weight along these paths grows, 
and normalization suppresses the mean intensity at the end. For $\eta^2<1$, 
the total weight stays finite and the mean intensity survives. At $\eta^2=1$, the 
exponential factors balance and the continuum mean falls slowly as $(\pi s)^{-1/2}$. 
A separate finite-chain argument gives the exact last-site limit in Eq.~\eqref{eq:end}. 

The mean stationary probability beyond a given Krylov position has an exact interpretation 
in classical diffusion. It equals the probability that a diffusing particle ever reaches 
a corresponding point against the mean tilt of a Brownian potential, averaged over the 
disorder [End Matter Eq.~\eqref{em:maximum}]. The Krylov endpoint threshold corresponds 
to the tilt at which the law of the maximum excursion changes~\cite{KawazuTanakaMaximum1993}.

\begin{figure}[t!]
 \centering
 \includegraphics[width=\linewidth]{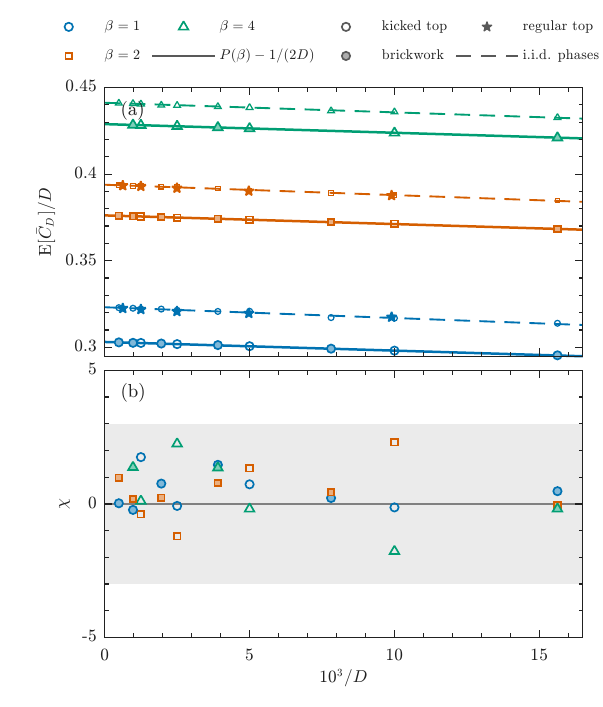}
 \caption{Mean Krylov spread for driven models. 
 (a) Kicked tops (open symbols) and circuits (filled symbols) for $\beta=1,2,4$ 
 approach the corresponding values of $P(\beta)$ at $1/D=0$. The solid lines 
 include the finite-size correction of $-1/(2D)$, as supported by the exact chain. 
 Small symbols and dashed lines give independent phases and their 
 fits $P_{\mathrm P}+b_{\mathrm P}/D$. Stars give the regular kicked top 
 with Gaussian initial states. 
 (b) Deviations $\chi$ of the chaotic-model data from the solid lines in units of 
 one standard error over 200 operators per size. The shaded band marks $|\chi|\leq3$.}
 \label{fig:A3}
\end{figure}

\begin{figure*}[t!]
 \centering
 \includegraphics[width=\textwidth]{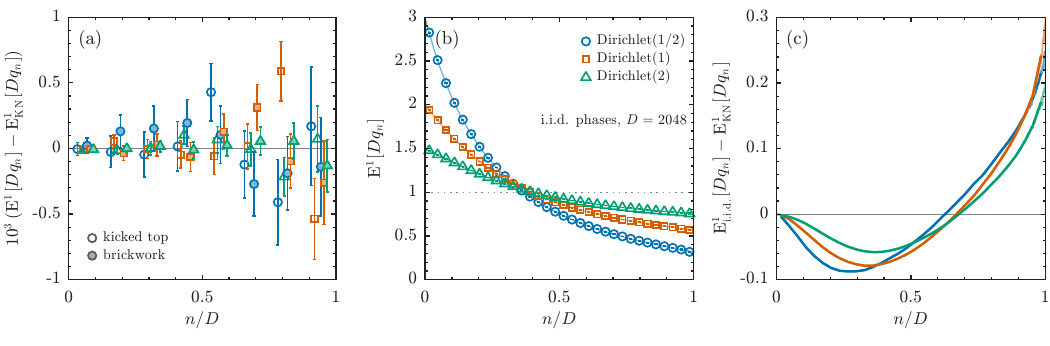}
 \caption{Effect of the spectrum at fixed overlap statistics. 
(a) Difference $\mathrm E^1[Dq_n]-\mathrm E^1_{\mathrm{KN}}[Dq_n]$
between the model and circular-ensemble mean occupations in units of
$1/D$ [Eq.~\eqref{eq:main1}]. Here, KN denotes the finite-$D$ reference
generated by the Killip--Nenciu construction [Eq.~\eqref{em:kn}],
using the same initial-weight statistics as the model
[Eq.~\eqref{em:weights}]. For each model, the difference is averaged
over eight bins in $n/D$ at the largest $D$ studied. The vertical scale 
is multiplied by $10^3$. Open and filled symbols denote tops and circuits. 
Errors give one standard error over operators. 
(b) Occupation for independent uniform phases at $D=2048$, with the 
weight distributions of the three symmetry classes [Dirichlet($\beta/2$) 
for $\beta=1,2,4$]. 
(c) Subtracting the corresponding circular-ensemble profile exposes 
a transfer of probability from the interior toward the far end. 
Its first moment accounts for the larger spread in Fig.~\ref{fig:A3}(a).}
 \label{fig:A4}
\end{figure*}

\paragraph{Tests on chaotic Floquet models.}
We tested the theory on a periodically kicked spin (the kicked top) and on a chain of 
qubits, repeating a fixed sequence of random two-qubit gates every period. The two 
models give mean adjacent-spacing ratios consistent with their corresponding circular 
ensembles (End Matter). Their mean spread and last-site occupation follow 
the finite-size predictions in all three symmetry classes 
[Figs.~\ref{fig:A3} and~\ref{fig:A2}(a)]. We resolve one endpoint excess 
of $(1.0\pm0.2)\%$ for the symplectic kicked top, which we trace to its nonuniversal 
spectral form factor at the first few periods (Supplemental Sec.~\ref{sm:sff}). 
At the largest size of each model, the eight-bin averaged occupations agree with 
the circular ensemble to within $6\times10^{-4}$ of the uniform share [Fig.~\ref{fig:A4}(a)].

Drawing an independent Haar-random unitary at every step gives a different
benchmark~\cite{Sahu}. Orthogonalizing the successive states again gives an 
ordered basis. Once the basis is complete, each subsequent state is Haar-distributed 
and independent of it, so that it has mean occupation $1/D$ in every direction and mean 
position $(D-1)/2$. Along the Floquet orbit of a single fixed unitary, the magnitudes 
of the overlaps with its eigenvectors are conserved. For a Haar-distributed drive, 
the ensemble-averaged spread satisfies $\mathrm E[\bar C_D]/D\to P(2) \approx 0.3762$.

\paragraph{What the quasienergy correlations contribute.}
We compare circular-ensemble spectra with independent uniform phases and draw 
the initial weights from the same law in both cases. Each realization uses 
a fixed Floquet operator. Its spectrum and the return probability have the same 
distributions in the two ensembles. The occupations farther down the chain distinguish them.
Independent phases shift probability from the interior, around $n/D=0.3$--$0.4$, toward 
the far end [Fig.~\ref{fig:A4}(b,c)]. All three weight laws show an increase
[Fig.~\ref{fig:A1.0}(c) and Supplemental Sec.~\ref{sm:numerics}]. 
For the sizes studied, the mean last-site occupation with independent phases is 
a substantial fraction of $1/D$ for each weight law [Fig.~\ref{fig:A2}(a)]. Independent 
phases remove correlations at all separations, so that this comparison shows 
their combined effect on the stationary profile.

Late-time Krylov complexity can be larger in an integrable regime than in
a chaotic one, depending on the initial operator or
state~\cite{PhysRevE.107.024217,p1}. This dependence has been traced to
how the initial condition distributes its weight over the
eigenstates~\cite{p1}. Level correlations 
in an unfolded spectrum also affect the finite-time peak of the spread~\cite{unfolded2}. 
Our control isolates their contribution to the stationary spread at fixed overlap 
statistics. For initial states matched to the symmetry class, the value $P(\beta)$ 
supplies a quantitative test of the circular-ensemble prediction.

The regular (integrable) kicked top provides a complementary example. 
With Gaussian random initial states, its spread follows the independent-phase 
benchmark [stars in Fig.~\ref{fig:A3}(a)]. Earlier work found a spread of 
order $\sqrt D$ for the same top with spin-coherent initial states in a 
reflection-symmetry sector~\cite{streamed}. Integrable dynamics can support 
markedly different late-time spreads, depending on the initial state and the symmetry sector.

\paragraph{Observability.} 
The chain described above can be implemented with two alternating layers
of beam splitters. With the independent reflection laws of End Matter
Eq.~\eqref{em:kn}, the network realizes the circular-$\beta$ quasienergies
and Dirichlet$(\beta/2)$ initial weights for any
$\beta>0$~\cite{Killip,TruninHuse2025}. The same two layers are repeated
each period. For a walker launched at the first site, the time-averaged
intensity at each output gives the stationary Krylov occupation $\bar p_n$.

In a chaotic drive, the quasienergies and overlaps can also be reconstructed from
the complex return amplitude $\langle\psi_0|U^t|\psi_0\rangle=\sum_a w_az_a^t$.
End Matter Eq.~\eqref{em:last} then gives $\bar p_{D-1}$ from these spectral
data. Resolving weak lines in the return signal matters because small overlaps
can strongly affect the last-site occupation. In principle, a size-scaling test
can use one realization at each chain length. The slow convergence with $D$
[Fig.~\ref{fig:A2}(a)] calls for a range of lengths (Supplemental
Sec.~\ref{sm:observables}).

The same scattering law arises in energy space. In a single-channel ring threaded 
by a linearly increasing flux, Zener transitions at avoided crossings couple 
neighboring adiabatic levels in two alternating 
layers~\cite{GefenThouless1987,BlatterBrowne1988}. 
Because level velocities grow with energy, the mean probability of Bragg 
reflection at the $k$th crossing falls as $\eta^2/(k+1)$. For Gaussian 
white-noise disorder, its law is exactly that of our chain read from its far end, 
with $\eta^2$ tuned by the electromotive force. If the reflection phases 
are independent and uniform, the ground-level spectral measure is continuous 
with dimension $1-\eta^2$ for $\eta^2<1$ and atomic for $\eta^2>1$~\cite{Simon_2005_2,brow3} 
(Supplemental Sec.~\ref{sm:energy}).

\paragraph{Conclusion.} The stationary occupation of the Krylov chain retains 
a measurable memory of the initial state even when local observables have relaxed 
to their infinite-temperature values. Long-time return measures how unevenly 
the initial state populates the Floquet eigenmodes. At fixed-overlap statistics, 
repelling and independent quasienergies yield the same return distribution but 
different occupations farther down the chain. Extensive spread and depletion 
at the last site can coexist. For example, the mean position grows in proportion 
to $D$ in both the unitary and symplectic classes, while the last site carries 
a vanishing fraction of $1/D$ in the unitary class and half of that share 
in the symplectic class.

Varying the initial preparation at fixed drive would test how the smallest eigenmode 
weights affect the endpoint. At fixed mean reflection law, one can vary the correlations 
between scatterers and test whether the threshold shifts. Energy-space localization 
has been studied in the kicked rotor and in interacting 
driven systems~\cite{FishmanGrempelPrange1982,DAlessioPolkovnikov2013}. 
In a flux-driven ring with point scattering, isolated avoided crossings yield 
inverse-distance reflection along the ladder of ordered 
energy levels~\cite{GefenThouless1987,BlatterBrowne1988}. 
This gives a physical setting in which to ask how the phases accumulated 
between crossings affect propagation up an energy ladder.

Superconducting-qubit experiments on the kicked top have measured 
local equilibration~\cite{NeillEtAl2016}. In a given Floquet system, interferometry 
of the return amplitude gives access to the spectral weights and quasienergies 
that determine the Krylov occupations. A programmable network of beam splitters, 
or a qubit circuit in its one-excitation sector, offers a direct 
test~\cite{Killip,TruninHuse2025}. Set the reflectivities and phases using 
the laws of End Matter Sec.~\ref{em:ensemble}. Launch an excitation at the 
first site, repeat the same two layers each period, and measure the time-averaged 
population at the far end in units of $1/D$. Varying the chain length and 
the reflection strength can test Eq.~\eqref{eq:end} and the finite-size curves in Fig.~\ref{fig:A2}(a).

\paragraph{Acknowledgments.} EMG thanks Matanya Wiener and Kamal Das for their 
help with the Weizmann Institute's EXAscale Cluster (WEXAC).
The work was supported by the Deutsche Forschungsgemeinschaft (DFG,
German Research Foundation) through grants SH 81/8-1 and GO 1405/7-1. 
YG also acknowledges support from the 
National Science Foundation (NSF)--Binational 
Science Foundation (BSF) through grant 2023666.

\paragraph{Data availability.} The data and code supporting this study 
are not publicly available but are available from the corresponding author 
on reasonable request.

\bibliography{refs-floquet.bib}

\section*{End Matter}

\section{Ensembles and scattering chains}\label{em:ensemble}
The forward Krylov basis is obtained by Arnoldi iteration, which implements
Gram--Schmidt orthogonalization of the ordered sequence
$\ket{\psi_0},U\ket{\psi_0},\ldots$~\cite{flo1,flo2,PhysRevE.108.054222,PhysRevE.110.034201,PhysRevB.110.155109}.
If $P_a$ is the projector onto the eigenspace of quasienergy $z_a=e^{i\theta_a}$ 
with eigenphase $\theta_a$, the 
span of the vectors $U^t\ket{\psi_0}$ intersects each occupied eigenspace in the 
line spanned by $\ket{z_a}=P_a\ket{\psi_0}/\sqrt{w_a}$, with $w_a=\|P_a\psi_0\|^2$. 
The cyclic dimension counts distinct eigenphases with nonzero projection of the 
initial state~\cite{_indrak_2024}. Thus, Eq.~\eqref{eq:cesaro} holds on the cyclic 
space also for Kramers-degenerate spectra. Since $\sum_a|\varphi_a(n)|^2=1$ on this 
space, Eq.~\eqref{eq:cesaro} also gives $\bar p_n\geq\min_a w_a>0$ at finite $D$. 
The corresponding overlap decomposition appears in
Refs.~\cite{suppression,competitor1,balasubramanian_variations_2026,Pain_2026}. 

Write $c=\beta/2$. The joint law used in the Letter consists of phases with 
density proportional to $\prod_{a<b}|e^{i\theta_a}-e^{i\theta_b}|^\beta$ relative 
to independent uniform phases, together with weights independent of the spectrum,
\begin{equation}\label{em:weights}
 w_a=\frac{G_a}{\sum_b G_b},\quad
 G_a\ \text{independent with law } \mathrm{Gamma}(c,1).
\end{equation}
This normalized gamma law is the symmetric Dirichlet distribution. 
At $c=1/2,1,2$, the variables $G_a$ are proportional to sums of squares 
of one, two, or four independent real Gaussian components, respectively. 
Direct averaging of $\bar p_0=\sum_a w_a^2$ gives
\begin{equation}\label{em:origin}
 D\,\mathrm E[\bar p_0]=\frac{D(c+1)}{Dc+1}.
\end{equation}

Gram--Schmidt in the spectral representation constructs polynomials $\phi_n$ 
orthonormal for $\sum_a w_a\delta_{z_a}$, with $\ket{K_n}=\phi_n(U)\ket{\psi_0}$. 
The intensity profile of mode $a$ is $q_n(z_a)=|\varphi_a(n)|^2=w_a|\phi_n(z_a)|^2$. 
One step of the recursion for $0\leq k\leq D-2$ reads~\cite{Simon2005Orthogonal,Simon_2005_2}
\begin{equation}\label{em:szego}
 \rho_k\phi_{k+1}(z)=z\phi_k(z)-\alpha_k^*\widetilde\phi_k(z),
 \quad \rho_k=\sqrt{1-|\alpha_k|^2},
\end{equation}
where $\widetilde\phi_k(z)=z^k \phi_k^*(1/z^*)$. 
The complex numbers $\alpha_k$, called Verblunsky coefficients, are reflection amplitudes:
each pair $(\alpha_k,\rho_k)$ specifies a lossless two-channel scatterer with reflection
and transmission probabilities $|\alpha_k|^2$ and $\rho_k^2$. 
The Killip--Nenciu construction~\cite{Killip} gives the phase and weight laws above 
by drawing these amplitudes independently with uniform angles and
\begin{equation}\label{em:kn}
 |\alpha_k|^2\sim{\rm Beta}(1,c_k),\qquad
 c_k=c(D-1-k),\quad 0\le k\le D-2 .
\end{equation}
The last coefficient has unit modulus and a uniform phase. 
At a distance $r=D-1-k$ from the end, the mean reflection probability is $1/(1+cr)$. 
The coefficient of the inverse-distance law is $\eta^2=2/\beta$, and the transition 
of Eq.~\eqref{eq:end} occurs at $\eta^2=1$. Supplemental Sec.~\ref{ssec:palm} relates 
this marginal decay to other random chains.

\section{The sampled eigenmode}\label{em:palm}
First draw the drive. Then draw one of its eigenmodes with probability $w_a$ and 
rotate all eigenphases by a common angle so that the sampled eigenvalue lies at $z=1$. 
The law of the environment seen from this sampled mode is called 
the Palm law~\cite{r1,r3,Valk__2023}; Supplemental Sec.~\ref{ssec:palm} gives 
its construction. Let $\mathrm E^1$ denote its expectation, and $q_n$ the 
normalized profile of the sampled mode. Equation~\eqref{eq:cesaro} gives
\begin{equation}\label{eq:main1}
 \mathrm E[\bar p_n]=\mathrm E^1[q_n],\qquad
 q_n=\frac{M_n}{\sum_{m=0}^{D-1}M_m},\quad M_n=|\phi_n(1)|^2 .
\end{equation}

By Eq.~\eqref{em:szego}, the ratios $X_k=M_{k+1}/M_k$, with $M_0=1$, 
are
\begin{equation}\label{em:Xk}
 X_k=\frac{|1-\widetilde\alpha_k|^2}{1-|\widetilde\alpha_k|^2},\qquad
 \widetilde\alpha_k=\alpha_k^*e^{-2i\arg\phi_k(1)}.
\end{equation}
Before weighting by the sampled mode, the conditional law 
of $\widetilde\alpha_k$ given the preceding sites has a uniform angle 
and the radial distribution in Eq.~\eqref{em:kn}. Thus, the $X_k$ are independent. 
To find the law of $X_k$, set $(1+\widetilde\alpha_k)/(1-\widetilde\alpha_k)=u+iv$. 
Then, $u=1/X_k>0$, and the joint density in the right half-plane is 
proportional to $u^{c_k-1}[(1+u)^2+v^2]^{-c_k-1}$. Integration over $v$ gives 
a marginal proportional to $u^{c_k-1}(1+u)^{-2c_k-1}$. Hence, $1/X_k$ has 
the beta-prime law with parameters $(c_k,c_k+1)$ and $X_k$ has the 
parameters in reverse order. The Palm factor $\prod_{k=0}^{D-2}X_k^{-1}$ 
tilts each marginal by $X_k^{-1}$. Consequently,
\begin{equation}\label{em:ratios}
 X_k\sim{\rm BetaPrime}(c_k,c_k+1),\qquad
 M_n=\prod_{k<n}X_k.
\end{equation}
In particular, $\mathrm E^1[X_k]=1$. A beta-prime variable with 
parameters $(a,b)$ is the ratio of independent unit-scale gamma 
variables of shapes $a$ and $b$.
The resulting mean occupation is strictly decreasing at every finite $D>1$. 
Condition on all ratios except $X_n$. For some fixed $A,B>0$, 
the conditional difference takes the form
\[
 q_{n+1}-q_n=M_n\,\frac{X_n-1}{A+BX_n}.
\]
Since $X_n$ has mean one and $1/(A+BX_n)$ decreases strictly with $X_n$, 
the conditional mean is negative. In particular, $\mathrm E[\bar C_D]<(D-1)/2$.

\section{Brownian limit and closed form}\label{em:moments}
The logarithm of $M_n$ is a sum of independent increments. 
For the beta-prime law, $\mathrm E^1[\ln X_k]=-1/c_k$ 
and $\mathrm{Var}(\ln X_k)=\psi_1(c_k)+\psi_1(c_k+1)=2/c_k+O(c_k^{-2})$, 
where $\psi_1$ is the trigamma function. 
On the scale $s=\ln[D/(D-n)]$, the accumulated mean and variance 
tend to $-2s/\beta$ and $4s/\beta$. On every bounded interval of $s$, 
the sum of the centered fourth moments is $O(D^{-1})$. The functional 
central limit theorem for independent increments~\cite{billingsley1999convergence} 
gives $\ln M_n\Rightarrow 2B(s)/\sqrt{\beta}-2s/\beta$ there. This explains 
the Brownian motion in Eq.~\eqref{eq:gibbs}~\cite{Valk__2023}. 
In the normalized profile, the site count contributes $\dd x=e^{-s}\dd s$. 
Consequently,
\begin{equation}\label{em:Z}
 \mathcal Z_\infty=\int_0^\infty
 \exp\!\left[\frac{2B(u)}{\sqrt\beta}
-\left(1+\frac{2}{\beta}\right)u\right] \dd u .
\end{equation}
The mean site profile converges to $g_\beta(x)=\mathrm E[\mathcal G_\beta(x)]$ 
pointwise for $0<x<1$ (Supplemental Sec.~\ref{sm:convergence}), 
and Eq.~\eqref{em:origin} covers $x=0$.

For evaluation of Fig.~\ref{fig:A1.0}, 
put $\kappa=(\beta-2)^2/(8\beta)$, $\mu=1+\beta/2$, and $s=-\ln(1-x)$. 
The explicit density is
\begin{equation}\label{eq:density}
 g_\beta(x)\!=\!\mu\,\sqrt{\frac{\beta}{8\pi s^3}}\,e^{-\kappa s}\!\int_0^\infty\!\! r\,e^{-\beta r^2/(8s)-(\mu/2+1)r}
  \mathcal I_\mu(e^{-r})\,\dd r ,
\end{equation}
where $\mathcal I_\mu(w)=\int_w^1t^{-2}(1+w-t)^{-\mu-1}\dd t$. For $p\geq0$, 
averaging the ratio of two integrals along the same Brownian path 
gives $\int_0^1(1-x)^p g_\beta(x)\ \dd x=F(\mu,\beta p)$. 
Setting $p=1$ and using Eq.~\eqref{eq:universal} gives Eq.~\eqref{eq:F}
of the main text. Integer values of $p$ give all integer moments 
(Supplemental Secs.~\ref{sm:plateau} and~\ref{sm:convergence}). 
The coefficient $P(\beta)$ is positive, and it is strictly less than $1/2$ because
$g_\beta$ is a decreasing, normalized density with $g_\beta(0)>1$. 
The asymptotes of the mean position are $P(\beta)\simeq(\pi^2/12)\beta$ 
at small $\beta$ and $P(\beta)\simeq 1/2-1/(3\beta)$ at large $\beta$. The density per 
unit logarithmic distance,
$e^{-s}\mathcal G_\beta(1-e^{-s})$, is the Boltzmann weight on the half-line in the tilted
Brownian potential $-2B(s)/\sqrt{\beta}+(1+2/\beta)s$~\cite{yor,BOUCHAUD1990285}.

\section{The endpoint}\label{em:endpoint}
Write $\eta^2=2/\beta$ and let $M(s)=\exp[\sqrt2\eta B(s)-\eta^2 s]$ be the continuum limit of $M_n$. 
Almost surely, $B(s)/s\to0$. 
Hence, $\ln\mathcal G_\beta(1-e^{-s})/s\to-\eta^2$ even though the mean endpoint 
density can be positive. The exponential $M(s)$ has expectation one. 
By the Girsanov theorem, weighting the Brownian path by $M(s)$ shifts $B(u)$ 
by $\sqrt2\eta\min(u,s)$. This Gaussian change of mean 
in $\mathrm E[M(s)/\mathcal Z_\infty]$ gives
\begin{align}
 &g_\beta(1-e^{-s})=\mathrm E\!\left[Z_s^{-1}\right], \label{em:tilt}\\
 &Z_s=\int_0^\infty
  e^{\sqrt2\eta B(u)-(1+\eta^2)u+2\eta^2\min(u,s)}\,\dd u . \nonumber
\end{align}

At fixed $u<s$, the tilt changes the systematic growth rate of $\ln M(u)$ 
from $-\eta^2$ to $+\eta^2$. The factor $e^{-u}$ from the site count supplies 
the competing rate $-1$. As $s\to\infty$, $Z_s$ increases 
to $\int_0^\infty e^{\sqrt2\eta B(u)-(1-\eta^2)u}\dd u$. For $\eta^2\geq1$, this 
integral diverges almost surely, including at zero drift by Brownian recurrence. 
For $\eta^2<1$, Brownian scaling and Dufresne's identity~\cite{Dufresne01011990,yor} 
give the limiting integral the law $\beta/(2\gamma_{\beta/2-1})$, 
where $\gamma_\nu\sim{\rm Gamma}(\nu,1)$. Since $Z_s^{-1}\leq Z_0^{-1}$ 
and $\mathrm E[Z_0^{-1}]=1+\eta^2$, dominated convergence 
gives $g_\beta(1^-)=(1-\eta^2)_+=(1-2/\beta)_+$. The more detailed continuum edge 
laws, derived from Eq.~\eqref{eq:density} in Supplemental Sec.~\ref{sm:edge}, are
\begin{equation}\label{em:continuumedge}
 g_\beta(1-e^{-s})\sim
 \begin{cases}
 C_\beta e^{-\kappa s}s^{-3/2},&\beta<2,\\
 (\pi s)^{-1/2},&\beta=2,\\
 1-2/\beta,&\beta>2 ,
 \end{cases}
 \qquad s\to\infty ,
\end{equation}
where $C_\beta>0$. 

The last few ratios in Eq.~\eqref{em:ratios} have $c_k=O(1)$ at all $D$. The 
continuum limit at fixed $x<1$ therefore requires a separate argument at $n=D-1$.
Monotonicity of the finite-chain mean gives $\limsup_{D\to\infty} y(D)\leq g_\beta(x)$ 
for every $0<x<1$. Letting $x\to 1^-$ proves Eq.~\eqref{eq:end} when $\beta\leq2$ 
and supplies the upper bound for $\beta>2$. 
For the lower bound, put $L(z)=\prod_a(z-z_a)$ 
and $b_a=|L'(z_a)|^{-2}/\sum_j|L'(z_j)|^{-2}$. 
Orthogonality of the last Krylov vector to $U^k\ket{\psi_0}$ for $0\leq k<D-1$, 
together with the interpolation identity $\sum_a z_a^k/L'(z_a)=0$, gives
\begin{equation}\label{em:last}
 q_{D-1}(z_a)=\frac{b_a/w_a}{\sum_j b_j/w_j},\qquad
 \bar p_{D-1}=\frac{1}{\sum_j b_j/w_j}.
\end{equation}
Here, $b_a$ depends only on the phases. At $c=\beta/2>1$, 
the Dirichlet weights obey $\mathrm E[1/w_a]=(Dc-1)/(c-1)$. 
Conditional Jensen's inequality at fixed phases gives
\[
 y(D)\geq\frac{D(c-1)}{Dc-1}
\]
at every finite $D$. This matches the continuum upper bound as $D\to\infty$ 
and proves the remaining case of Eq.~\eqref{eq:end}. 
The approach rates are analyzed in Supplemental Sec.~\ref{sm:lattice}.

The cumulative Krylov occupation has a direct first-passage interpretation. 
Let $W(u)$ be a two-sided standard Brownian motion, with independent paths 
on either side of the origin. Consider the diffusion started at zero in the 
potential $V(u)=W(u)+hu$, with generator 
$(1/2)e^{V(u)}(\dd /\dd u)[e^{-V(u)}\dd / \dd u]$. 
For $h>0$, it is left-biased. With $\xi_h$ the furthest point it ever 
reaches on the right and both the thermal motion and the disorder included in $\Pr$, 
\begin{equation}\label{em:maximum}
 \int_{1-e^{-s}}^1 g_\beta(x)\,\dd x
  =\Pr\!\left(\xi_h>\frac{4s}{\beta}\right),\qquad
 h=\frac{\beta+2}{4}.
\end{equation}
Kawazu and Tanaka obtained the tail of $\xi_h$ in three regimes, 
which meet at ${h=1}$, i.e., at $\beta=2$~\cite{KawazuTanakaMaximum1993}. 
Supplemental Sec.~\ref{sm:passage} proves Eq.~\eqref{em:maximum} and reproduces Eq.~\eqref{em:continuumedge}.

\section{Numerical tests}\label{em:profiles}
For Fig.~\ref{fig:A1.0}, we used 200 independently drawn circular spectra 
at $D=2048$ for each $\beta=1/2,1,4/3,2,3,4,6,8$, with 50 independent weight 
vectors per spectrum. The reference in panel (b) is $g_\beta(n/D)$ for $n\geq1$ 
and the finite-$D$ value in Eq.~\eqref{em:origin} at $n=0$. 
Each point with $n\leq D-6$ lies within three standard errors of its reference.

The kicked tops~\cite{haake1987classical,kicked1} and brickwork circuits~\cite{C1}, 
their parameters and sizes, and the independent-phase control are specified 
in Supplemental Sec.~\ref{sm:models}. The mean ratio of adjacent quasienergy 
spacings lies within two standard errors of the large-$D$ circular-ensemble value 
at each of the 27 model-size points (Supplemental Fig.~\ref{fig:rbar}). 
For the plateau, the average at each of the 27 points first runs 
over 100--400 initial states per operator and symmetry sector and 
then over 200 operators. For the finite-size benchmark $P(\beta)-1/(2D)$, 25 
of the 27 plateau residuals lie within two operator-level standard errors, 
all lie within $2.31$, and their sum of squares in these units is $29.3$, against 
an expected value of 27. Figure~\ref{fig:A4}(a) compares eight bins of 
the profile at the largest size of each model against circular spectra with 
matched weights. In these bins, the maximum absolute difference is $6\times10^{-4}$ 
in units of the uniform occupation.

\begin{widetext}

\setcounter{section}{0} 
\renewcommand{\thesection}{S\arabic{section}}

\renewcommand{\thesubsection}{\thesection.\arabic{subsection}}
\makeatletter\renewcommand{\p@subsection}{}\makeatother


\setcounter{equation}{0}
\setcounter{figure}{0}
\setcounter{table}{0}
\setcounter{page}{1}
\setcounter{section}{0}

\renewcommand{\theequation}{S\arabic{equation}}
\renewcommand{\thefigure}{S\arabic{figure}}
\renewcommand{\thetable}{S\arabic{table}}
\renewcommand{\thepage}{S\arabic{page}}
\renewcommand*{\theHsection}{S.\arabic{section}}
\renewcommand*{\theHequation}{S.\arabic{equation}}
\renewcommand*{\theHfigure}{S.\arabic{figure}}
\renewcommand*{\theHtable}{S.\arabic{table}}
\newcommand{\blk}{\text{bulk}}

\section*{Supplemental Material}

\section{Krylov construction}
\subsection{Setup}\label{ssec:setup}
Henceforth, as in the Letter, unless stated otherwise, the spectrum of every $U \in \mathbb{C}^{D\times D}$ is simple with $\{ z_a = e^{i\theta_a}\}_{a=1}^D$, $\theta_1< \dotsi < \theta_D$, $\theta_a \in [0,2\pi)$. Within each realization, the initial vector (the seed) $\ket{\psi_0}$ is fixed in the ambient basis of the Hilbert space $\mathcal{H} = \mathbb{C}^D$, such that it has non-zero overlap with all the eigenstates of $U$, $\ket{z_a}$, i.e., $w_a \coloneqq \abs{\bra{z_a}\ket{\psi_0}}^2>0$ for every $a$. Consequently, the Krylov subspace of the stroboscopic evolution is $\mathcal{K}(U,\psi_0) = \mathrm{span}\{U^t\ket{\psi_0}\, \vert \, t\in \mathbb{Z}_{\geq 0}\} = \mathcal{H}$, so $\mathrm{\dim}\, \mathcal{K}(U,\psi_0) = D$.
For the Kramers-degenerate spectra of the symplectic class, $D$ counts the distinct eigenphases and $\ket{z_a}$ is the normalized projection of the seed onto the corresponding eigenspace, as in End Matter Sec.~\ref{em:ensemble}; the statements below then hold on the cyclic subspace. Once $U$ becomes random, a single trajectory, or realization of the evolution protocol, consists of keeping $U$ fixed and evolving $\ket{\psi_0}$. Then, the infinite-time average of the Krylov spread (the plateau) is calculated and stored, and this is repeated many times to perform the ensemble average of the plateaus. 

For the ensemble average, we use the matched spectral measure of the circular $\beta$-ensemble (C$\beta$E) in the Killip--Nenciu (KN) construction. Its spectral weights have the law
\begin{equation}\label{eq:dir}
  (w_1,\ldots, w_D) \sim \mathrm{Dirichlet}(\beta/2,\ldots, \beta/2),
\end{equation}
where $\beta>0$ is the Dyson index, and are independent of the eigenphases of $U$, which have the joint law
\begin{equation}
  \mathrm{P}_{D,\beta}(\dd \theta_1, \ldots, \dd \theta_D) \propto \prod_{1\leq j < k \leq D} \abs{e^{i \theta_j} - e^{i\theta_k}}^\beta \prod_{j=1}^D \dfrac{\dd \theta_j}{2\pi}. 
\end{equation}
In the notation of the End Matter, the Dirichlet parameter is $c=\beta/2$. The weights depend on the initial state, not only on the drive: an independent complex Haar seed in an orthogonal eigenbasis has $c=1$, not $c=1/2$. Such a seed lowers $D\,\mathrm E[\bar p_0]$ at large $D$ from 3 to 2 [Eq.~\eqref{em:origin}] and lies outside the matched law $c=\beta/2$ assumed in Eqs.~\eqref{eq:F} and~\eqref{eq:density}.

\subsection{Orthonormal polynomials on the circle and the Krylov construction}
Here, we briefly discuss the tools of orthonormal polynomials on the circle (OPUC) we implemented to obtain the results in the Letter. For an authoritative treatment of OPUC, see Refs.~\cite{Simon2005Orthogonal,Simon_2005_2}; for their application to the Krylov construction, see Refs.~\cite{flo1,flo2,streamed}.

Denote the unit open disk and unit circle on $\mathbb C$ by $\mathbb{D} = \{ z\in \mathbb{C} : \abs{z}<1 \}$ and $\mathbb{T} =\partial \mathbb{D}$. The pair $(U,\psi_0)$ defines a probability measure on $\mathbb{T}$ with finite support (called trivial in the OPUC literature) 
\begin{equation}\label{eq:measure}
\mu = \sum_{a=1}^D w_a \delta_{z_a},  
\end{equation}
supported on $D$ atoms $\{z_a\}$. The inner product of two functions $f,g\in L^2(\mathbb T, \mu)$ is
\begin{equation}
  \langle f, g\rangle_\mu \coloneqq \int_{\mathbb T}f^*(z)g(z) \mu(\dd z) = \sum_{a=1}^D w_a f^*(z_a)g(z_a).
\end{equation}
The orthogonal monic polynomials of $\mu$ are $\{\Phi_n(z)\}_{n=0}^{D-1}$, $\langle \Phi_n, \Phi_m\rangle_\mu = \norm{\Phi_n}^2\delta_{nm}$. With $\Pi_{n-1}$ the orthogonal projector onto $\mathrm{span}\{1,z,\ldots,z^{n-1}\}$ and $\Pi_{-1}=0$, the monic polynomials $\Phi_n=(1-\Pi_{n-1})z^n$ have the form $\Phi_n(z) = z^n + r_{n-1}(z)$, where $r_{n-1}(z)$ is a polynomial of degree at most $n-1$. The orthonormal polynomials of $\mu$ are just
\begin{equation}
  \phi_n \coloneqq \dfrac{\Phi_n}{\norm{\Phi_n}} = \kappa_n z^n + \kappa_n r_{n-1}, \quad \kappa_n^2 \coloneqq \norm{\Phi_n}^{-2}. 
\end{equation}
The orthogonal and orthonormal polynomials obey the Szeg\H{o} recurrences
\begin{equation}\label{eq:recurrences}
  \Phi_{n+1}(z) = z\Phi_n(z) - \alpha_n^* \tilde\Phi_n(z), \quad  \phi_{n+1}(z) = \dfrac{1}{\rho_n}\left[ z \phi_n(z) - \alpha_n^* \Tilde{\phi}_n(z) \right], \quad n = 0, \ldots,D-2,
\end{equation}
where $\alpha_0,\ldots, \alpha_{D-1}$ are complex parameters called the \emph{Verblunsky coefficients} (VC), with 
\begin{equation}
  \alpha_{\text{bulk}} \coloneqq (\alpha_0, \ldots, \alpha_{D-2}) \in \mathbb{D}^{D-1}, \quad \alpha_{D-1} \in \mathbb{ T},
\end{equation}
the \emph{bulk} and \emph{edge} coefficients, respectively. Whenever convenient, we use the vector of $D$ VC, $\alpha = (\alpha_{\text{bulk}}, \alpha_{D-1})$; $\rho_k \coloneqq \sqrt{1-\abs{\alpha_k}^2}$, and for any polynomial $P_n$ of degree $n$, $\tilde P_n(z) \coloneqq z^n P_n(1/z^*)^*$ is the reversed polynomial. The VC are further related to the two types of polynomials via
\begin{equation}\label{eq:monic_VC}
  \norm{\Phi_{n+1}}^2 = \prod_{j=0}^{n}(1-\abs{\alpha_j}^2), \quad \alpha_n = -\Phi_{n+1}^*(0), \quad \dfrac{\abs{\phi_{n+1}(1)}^2}{\abs{\phi_n(1)}^2} = \dfrac{\abs{1-\widetilde\alpha_n}^2}{1-\abs{\alpha_n}^2}, 
\end{equation}
with the rotated coefficient $\widetilde\alpha_n=\alpha_n^*\exp[-2i\arg\phi_n(1)]$ of Eq.~\eqref{em:Xk}. From Eq.~\eqref{eq:recurrences}, one can see that, as a function of the VC,
\begin{equation}\label{eq:phi_vc}
  \phi_n = \phi_n(\alpha_0, \ldots, \alpha_{n-1})
\end{equation}
This fact is particularly important for simplifying the average of the mean Krylov probabilities.

To close this section, the reason for introducing the OPUC in the Krylov construction~\cite{flo2} is that the ordered list $\{U^t\}_{t=0}^{D-1}$ is mapped to $\{z^t \}_{t=0}^{D-1}$, from which, via Gram-Schmidt, one obtains the orthonormal polynomials $\{\phi_n\}_{n=0}^{D-1}$. As we show below, this fact is directly related to the Krylov-Arnoldi basis~\cite{flo2}.

\subsection{Arnoldi and CMV bases}\label{sm:arnoldi-cmv}
There are two main ways to find the basis spanning $\mathcal{K}$, the Krylov-Arnoldi basis $\{\ket{K_n}\}$~\cite{PhysRevE.108.054222}, and the Cantero-Moral-Velázquez (CMV) basis $\{\ket{C_n}\}_{n=0}^{D-1}$~\cite{CMV,streamed}. The former consists of applying Gram-Schmidt to the set $\{U^t\ket{\psi_0}\}$, so $U$ is an upper-Hessenberg matrix in $\ket{K_n}$. First, one sets $\ket{K_0 } = \ket{\psi_0}$, and the remaining $D-1$ vectors are found via the Arnoldi iteration 
\begin{equation}\label{eq:yy4}
  h_{n,n-1}\ket{K_n} = U\ket{K_{n-1}} - \sum_{j=0}^{n-1}h_{j,n-1}\ket{K_j}, \quad n \geq 1, \quad  h_{j,k} \coloneqq \bra{K_j}U\ket{K_k}, \quad h_{j,-1} \coloneqq 0
\end{equation}
where $h_{j,k}$ are the Arnoldi coefficients. The iteration, naturally, ends at step $n = D$, where $h_{D,D-1} = 0$ and no $\ket{K_{D}}$ is produced. Alternatively, for $z \in \mathbb T$, under the mapping $U^t\mapsto z^t$, the Krylov-Arnoldi vectors are
\begin{equation}\label{eq:krylov-arnoldi}
 \ket{K_n} = \phi_n(U)\ket{\psi_0}  
\end{equation}
and the entries of $U$ in this basis are 
\begin{equation}\label{eq:yy16}
  U_{mn} \coloneqq \langle \phi_m, z\phi_n\rangle_\mu = \bra{K_m}U\ket{K_n} = \begin{cases}
    -\alpha_n^*\alpha_{m-1}\prod_{k=m}^{n-1}\rho_k, \quad &m<n+1, \\
    \rho_n, \quad &m= n+1, \\
    0, \quad &m>n+1,
  \end{cases}
\end{equation}
with $\alpha_{-1} \coloneqq -1$ as an auxiliary VC. This construction is known as the Geronimus-Gragg-Teplyaev (GGT) form of the Arnoldi iteration~\cite{Simon_2005_2}, and its connection to Eq.~\eqref{eq:yy4} is the following: the subdiagonal is $h_{n,n-1} = \rho_{n-1}$; the diagonal is $h_{nn} = -\alpha_n^*\alpha_{n-1}$; the first row is $h_{0n} = \alpha_n^*\prod_{l<n}\rho_l$; and the upper triangle is the $m<n+1$ case of Eq.~\eqref{eq:yy16}. 

In contrast to the Arnoldi iteration, the CMV basis~\cite{CMV} is found by applying Gram-Schmidt to the first $D$ vectors in the alternating sequence
\[
 \ket{\psi_0},\ U\ket{\psi_0},\ U^{-1}\ket{\psi_0},\ U^2\ket{\psi_0},\ U^{-2}\ket{\psi_0},\ldots .
\] In terms of the OPUC, the basis vectors are $\ket{C_n} \coloneqq \chi_n(U)\ket{\psi_0}$, where $\chi_n$ are the CMV Laurent polynomials~\cite{Simon2005Orthogonal}, $\chi_{2k-1} = z^{-(k-1)}\phi_{2k-1}$ and $\chi_{2k} = z^{-k}\tilde \phi_{2k}$.
In terms of the Krylov-Arnoldi vectors, 
\begin{equation}
  \ket{C_n} = U^{-\lfloor n/2\rfloor}\, \begin{cases}
    \ket{K_n}, \quad &n \, \text{odd}, \\
    \ket*{\tilde{K}_n}, \quad &n \, \text{even}, 
  \end{cases}
\end{equation}
where $\ket*{\tilde{K}_n} \coloneqq \tilde{\phi}_n(U)\ket{\psi_0}$~\footnote{Our $\ket{C_n}$ vectors are called $\ket{P_{2n}}$ and $\ket{P_{2n+1}}$ in Ref.~\cite{streamed}.}. On this basis, the Floquet operator reads
\begin{equation}\label{eq:CMV}
 \mathcal{U}_{mn} \coloneqq \bra{C_m}U\ket{C_n} = ( \mathcal{L}\mathcal{M} )_{mn}, \quad \mathcal{L} = \Theta_0 \oplus \Theta_{2}\oplus~\dotsi, \quad   \mathcal{M} = 1_1\oplus \Theta_1 \oplus \Theta_3 \oplus \dotsi,\quad \Theta_k \coloneqq \begin{pmatrix}
    \alpha_k^* & \rho_k \\
    \rho_k & -\alpha_k
  \end{pmatrix},
\end{equation}
and the product $\mathcal{U} = \mathcal{L}\mathcal{M}$ is five diagonal. 

The Arnoldi and CMV bases span the same Krylov subspace, but not the same nested subspaces, since $\chi_n$ contains negative powers of $z$. To treat both at once, let $\ket{Q_n} \coloneqq \hat{\phi}_n(U)\ket{\psi_0}$, with $\hat{\phi}_n \in \{\phi_n, \chi_n \}$, denote the $n$-th vector of either basis. Their associated Krylov amplitude is $\psi_n(t) \coloneqq \bra{Q_n}\ket{\psi_t}$, $t \in \mathbb{Z}_{\geq 0}$, and the evolved state and its Krylov spread complexity read
\begin{equation}\label{eq:ec0}
  \ket{\psi_t} = \sum_{n=0}^{D-1}\psi_n(t)\ket{Q_n}, \quad C(t) = \sum_{n=0}^{D-1}n\abs{\psi_n(t)}^2,
\end{equation}
respectively. Since the spread weights each basis vector by its index $n$, $C(t)$ depends on the chosen basis. Its infinite-time average, derived below, does not, because $\abs{\chi_n(z)}=\abs{\phi_n(z)}$ on $\mathbb T$. Expanding the initial state in the eigenbasis of $U$, with the phases of the eigenvectors fixed by $\bra{z_a}\ket{\psi_0} = \sqrt{w_a}$, the spread becomes
\begin{equation}\label{eq:ec2}
  C(t) = \sum_{n,a}n w_a\abs{\bra{Q_n}\ket{z_a}}^2 + \sum_{n, a\neq b}n\sqrt{w_a w_b}z_a^t z_b^{-t}\bra{Q_n}\ket{z_a}\bra{z_b}\ket{Q_n}. 
\end{equation}
The second sum consists of oscillating terms of the form $(z_az_b^*)^t$, and since the spectrum of $U$ is simple, each of those terms has vanishing Ces\`aro mean, $\lim_{t \to \infty}t^{-1}\sum_{\tau=0}^{t-1}(z_a z_b^*)^\tau = 0.$ Hence, the infinite-time average of $C(t)$ exists and equals the Ces\`aro mean of the sequence $C(\tau)$, 
\begin{equation}\label{eq:ec3}
  \bar C_D \coloneqq \lim_{t\to \infty}\dfrac{1}{t}\sum_{\tau = 0}^{t-1} C(\tau) =\sum_{n,a}n w_a\abs{\bra{Q_n}\ket{z_a}}^2.
\end{equation}
This is the diagonal-ensemble representation of the plateau. From now on, we use the Krylov-Arnoldi basis, $\ket{Q_n} = \ket{K_n}$. Defining the Krylov probability profiles and barycenters
\begin{equation}\label{eq:profiles}
q_n(z_a) \coloneqq \abs{\bra{K_n}\ket{z_a}}^2, \quad \bar n_a \coloneqq \sum_{n=0}^{D-1}n q_n(z_a),  
\end{equation}
respectively, gives us the mean Krylov probabilities and the spread in the following forms
\begin{equation}\label{eq:integrals}
  \bar p_n = \sum_{a=1}^D w_a q_n(z_a) = \int_{\mathbb T} q_n(z) \mu(\dd z), \quad \bar C_D = \sum_{n=0}^{D-1}n \bar p_n = \sum_{a=1}^Dw_a \bar n_a. 
\end{equation}
From the last form of the spread, and since $q_n(z_a)$ is doubly stochastic, the empirical covariance of the uniformly weighted index $a$, i.e., with probability $1/D$, of $w_a$ and $\bar n_a$ is
\begin{equation}
  \mathrm{Cov}_a(w_a,\bar n_a) = \dfrac{1}{D}\sum_a w_a \bar n_a - \left( \dfrac{1}{D}\sum_a w_a \right)\left(\dfrac{1}{D}\sum_a \bar n_a\right) = \dfrac{\bar C_D}{D} - \dfrac{1}{D}\dfrac{D-1}{2},
\end{equation}
from which
\begin{equation}\label{eq:mean_cov}
  \bar C_D = \frac{D-1}{2} + D\, \mathrm{Cov}_a(w_a,\bar n_a).
\end{equation}
The shift of the mean position from $(D-1)/2$ thus measures how the center of an eigenmode's intensity profile on the chain correlates with its weight at the source.

\subsection{Averaging the last Verblunsky coefficient and the Palm measure}\label{ssec:palm}
Here, we demonstrate how to obtain Eq.~\eqref{eq:main1} in more detail, which requires $U$ to be drawn from the C$\beta$E and the initial state to obey the conditions of Sec.~\ref{ssec:setup}. To do so, we must first realize that the profiles $q_n(z_a)$ depend explicitly only on the bulk VC. Then, when taking the ensemble average over the VC, it is convenient to average the edge coefficient to subsequently center the quantities at $z=1$. 

The Krylov profiles~\eqref{eq:profiles} can be cast in terms of the polynomials by writing the initial state in the $\ket{z_a}$-basis together with Eq.~\eqref{eq:krylov-arnoldi}, which gives $q_n(z_a) = w_a \abs{\phi_n(z_a)}^2$. Using here the Christoffel function $w_a^{-1} = \sum_{m=0}^{D-1}\abs{\phi_m(z_a)}^2$ yields
\begin{equation}\label{eq:profiles2}
  q_n(z_a;\alpha_{\text{bulk}}) = \dfrac{\abs{\phi_n(z_a)}^2}{\sum_{m=0}^{D-1}\abs{\phi_m(z_a)}^2},
\end{equation}
where we have made the dependence on the bulk VC explicit, as given by Eq.~\eqref{eq:phi_vc}. The dependence on the edge coefficient is implicit through the roots of the paraorthogonal polynomial $\Phi_D(z) = 0 = z\Phi_{D-1}(z) - \alpha_{D-1}^*\tilde{\Phi}_{D-1}(z)$, which are the eigenvalues of $U$. As we show below, rotating $\alpha_{D-1}$ while keeping $\alpha_{\text{bulk}}$ fixed moves the atoms $z_a$ along the circle, yet the curve $\theta \mapsto q_n(e^{i\theta};\alpha_{\text{bulk}})$ remains fixed. For the matched law of the phases and weights in Sec.~\ref{ssec:setup}, the KN theorem~\cite{Killip} gives independent, rotationally invariant VC with the laws
\begin{equation}\label{eq:verb}
  \abs{\alpha_k}^2 \sim \mathrm{Beta}(1,c_k), \quad c_k \coloneqq \dfrac{\beta}{2}(D-1-k), \quad 0 \leq k \leq D-2, \quad \alpha_{D-1} = e^{i\chi}, \quad \chi \sim \mathrm{Unif}[0,2\pi). 
\end{equation}
Hence, they have the joint KN measure
\begin{equation}\label{eq:laws}
  \mathrm{P}(\dd \alpha) = \mathrm{P}_{\text{bulk}}(\dd \alpha_\text{bulk})\otimes \mathrm{P}_{D-1}(\dd \alpha_{D-1}) = \prod_{k=0}^{D-2}\dfrac{c_k}{\pi}\left( 1- \abs{\alpha_k}^2\right)^{c_k-1 }\dd^2\alpha_k\, \otimes \dfrac{\dd \chi}{2\pi},
\end{equation}
where $\dd^2\alpha_k = \dd (\mathrm{Re}\,\alpha_k)\,\dd(\mathrm{Im}\,\alpha_k)$, and the integration is carried over $\mathbb{D}^{D-1}\times \mathbb{T}$. Since the spectral measure ``knows'' about the VC, we shall write it as $\mu_\alpha = \mu_{(\alpha_{\text{bulk}},e^{i\chi})}$. Using this notation in the time-averaged Krylov probabilities in their integral form [Eq.~\eqref{eq:integrals}], and then taking the ensemble average over the VC yields
\begin{equation}\label{eq:campbell}
  \mathrm{E}_\mathrm{P}[\bar p_n] = \int_{\mathbb D^{D-1} \times \mathbb T}\int_{\mathbb{T}}q_n(z; \alpha_{\text{bulk}})\mu_{\alpha}(\dd z)\mathrm{P}(\dd \alpha) \eqqcolon \mathrm{E}_\mathcal{C}[q_n(Z;\alpha_{\text{bulk}})], 
\end{equation}
where $\mathcal{C}(\dd \alpha, \dd z) \coloneqq \mu_\alpha(\dd z) \mathrm{P}(\dd \alpha)$ is the Campbell measure of the probability measure $\mu_\alpha$, and where $Z$ has been promoted to a random variable. Notice that, under Eq.~\eqref{eq:integrals}, $Z = z_a$ is drawn with probability $w_a$, making the interpretation of the plateau as a size-biased average as given in Eq.~\eqref{eq:integrals}. Under a slight abuse of notation, the VC on the last equality of Eq.~\eqref{eq:campbell} are random variables. 

Since the profiles $q_n$ are nonnegative, we can use Tonelli's theorem in Eq.~\eqref{eq:campbell}, keep the bulk VC fixed, and average the spectral measure over the edge coefficient, which is just the Bernstein-Szeg\H{o} (BS) average~\cite{Simon2005Orthogonal}
\begin{equation}
   \mathrm{E}_\chi[\mu_{(\alpha_\text{bulk}, e^{i\chi})}(\dd \theta)] = \int_0^{2\pi}\mu_{(\alpha_\text{bulk}, e^{i\chi})} \dfrac{\dd \chi}{2\pi} = \dfrac{\dd \theta}{2\pi\abs{\phi_{D-1}(e^{i\theta}; \alpha_\text{bulk})}^2}, \label{eq:bs}
\end{equation}
where $z = e^{i\theta}$. The right-hand side is just the BS measure with $\alpha = (\alpha_\text{bulk},0)$. Replacing this back into Eq.~\eqref{eq:campbell} gives
\begin{equation}\label{eq:inte1}
  \mathrm{E}_{\mathcal{C}}[q_n(Z;\alpha_\blk)] = \int_{\mathbb{D}^{D-1}}\!\left[ \int_0^{2\pi} q_n(e^{i\theta}; \alpha_\text{bulk}) \dfrac{\dd \theta}{2\pi\abs{\phi_{D-1}(e^{i\theta}; \alpha_\text{bulk})}^2} \right]\mathrm{P}_\blk(\dd \alpha_\text{bulk}).
\end{equation}
Averaging over a fixed VC is standard in the OPUC literature~\cite{Simon2005Orthogonal}. 

For convenience, let us explicitly state the dependence of the orthonormal polynomials on the VC as 
$\phi_n(\cdot; \alpha)$. These depend only on the bulk VC, but we write $\alpha$ for simplicity. 
For a fixed $\vartheta$, let $\alpha(\vartheta)$ denote the Verblunsky coefficients of the rotated 
measure $\mu_\vartheta = \sum_{a}w_a \delta_{e^{i\vartheta} z_a}$, 
and $\phi_n(\cdot; \alpha(\vartheta))$ its orthonormal polynomials. 
From the Szeg\H{o} recursion~\eqref{eq:recurrences} obeyed by the two families of polynomials, 
one gets $\alpha_n(\vartheta) = e^{-i(n+1)\vartheta}\alpha_n$ and 
$\abs{\phi_n(e^{i\vartheta}z; \alpha(\vartheta))} = \abs{\phi_n(z;\alpha)}$ 
for every $z \in \mathbb T$ and every realization of $\alpha$. 
The integrand of Eq.~\eqref{eq:inte1}
\begin{equation}
  F_n(\theta;\alpha_\text{bulk}) \coloneqq q_n(e^{i\theta};\alpha_\text{bulk})\abs{\phi_{D-1}(e^{i\theta}; \alpha_\text{bulk})}^{-2},
\end{equation}
which depends on $\alpha_\blk$ only through the moduli $\abs{\phi_n}$, satisfies the pathwise 
identity $F_n(\theta+\vartheta;\alpha_\blk(\vartheta)) = F_n(\theta;\alpha_\blk)$. 
Since the $\alpha$ in Eq.~\eqref{eq:laws} are independent and each has a rotation-invariant law,
$\alpha_\blk(\vartheta) \overset{d}{=} \alpha_\blk$. 
Therefore, 
\begin{equation}
  F_n(\theta + \vartheta; \alpha_\blk) \overset{d}{=} F_n(\theta + \vartheta; \alpha_\blk(\vartheta)) = F_n(\theta;\alpha_\blk)
\end{equation}
for every fixed $\vartheta$; in particular, $F_n(\theta; \alpha_\blk) \overset{d}{=} F_n(0;\alpha_\blk)$ 
for every fixed $\theta$. As $F_n \geq 0$, Tonelli's theorem allows us to exchange the expectation 
and the $\theta$-integral in Eq.~\eqref{eq:inte1}, giving
\begin{equation}
  \mathrm{E}_\mathrm{P}[\bar p_n] = \int_0^{2\pi}\mathrm{E}_\blk\!\left[ F_n(\theta;\alpha_\blk) \right]
  \dfrac{\dd\theta}{2\pi}=\int_{\mathbb{D}^{D-1}}q_n(1;\alpha_\text{bulk})
  \dfrac{\mathrm{P_\blk(\dd \alpha_\text{bulk})}}{\abs{\phi_{D-1}(1;\alpha_\text{bulk})}^2}.
\end{equation}
The same argument applied to $\abs{\phi_{D-1}}^{-2}$ alone shows that 
$\mathrm{E}_\blk\abs{\phi_{D-1}(e^{i\theta};\alpha_\text{bulk})}^{-2}$ does not depend on $\theta$. 
Since the BS measure has unit mass, $\int_0^{2\pi}\abs{\phi_{D-1}(e^{i\theta}; \alpha_\blk)}^{-2}\dd \theta/(2\pi) = 1$
for every $\alpha_\blk$. Taking the expectation then gives $\mathrm{E}_\blk\abs{\phi_{D-1}(e^{i\theta};
\alpha_\blk)}^{-2} = 1$ for every $\theta$. Thus, we can define the probability measure
\begin{equation}\label{eq:Palm_meas}
  \mathrm{P}^{1}(\dd \alpha_\text{bulk}) \coloneqq \dfrac{\mathrm{P_\blk(\dd \alpha_\text{bulk})}}{\abs{\phi_{D-1}(1;\alpha_\text{bulk})}^2},
\end{equation}
which is the $\alpha_\text{bulk}$-marginal of the Palm distribution at $z = 1$ of the random spectral measure $\mu$.
Disintegrating its Campbell measure, $\mathrm{P}(\dd \alpha)\mu_\alpha(\dd z) = \mathrm{P}^z(\dd \alpha) \nu(\dd z)$,
with mean measure $\nu(\dd z) = \dd \theta/(2\pi)$, gives the Palm distributions $\mathrm{P}^z$, and $\mathrm{P}^1$ 
is the one at $z = 1$. This proves Eq.~\eqref{eq:main1}. The same law (i.e., the coefficient law biased by 
$\abs{\phi_{D-1}(1)}^{-2}$) was obtained by a different route in Ref.~\cite{Valk__2023}, 
where it is used to construct the Palm measure of the $\mathrm{Sine}_\beta$ process. 

The independent laws in Eq.~\eqref{eq:verb} give the mean reflection probability $1/(1+cr)$ at a distance $r=D-1-k$ from the end, with $c=\beta/2$, and a root-mean-square reflection amplitude scaling as $r^{-1/2}$. This is the marginal decay studied for random unitary chains by Killip and Stoiciu~\cite{Killip1} and for a random potential in a one-dimensional Schr\"odinger operator by Kiselev, Last, and Simon~\cite{KiselevLastSimon1998}. In the circular model, the coefficient of the inverse-distance law is $\eta^2=\lim_{r\to\infty}\lim_{D\to\infty}r\,\mathrm E|\alpha_{D-1-r}|^2=2/\beta$. Inferring an endpoint law from $\eta$ in other models requires information about the statistics and correlations of their reflection amplitudes.

Reading the reflection laws in Eq.~\eqref{em:kn} from the last link toward the first gives the sequence of beta parameters $c,2c,3c,\ldots$. The corresponding half-line model has finite sections with the same eigenphase statistics as our chain and different spectral weights~\cite{brow3}. For $\beta>2$, its spectral measure on the eigenphase circle has exact Hausdorff dimension $1-2/\beta$~\cite{Simon_2005_2,brow3} and a representation in terms of Gaussian multiplicative chaos~\cite{Chhaibi_2025}. 

\paragraph{Remark: the endpoint and the reversed chain.} The value $1-2/\beta$ appears twice: as the limit of the last-site occupation in Eq.~\eqref{eq:end} and as the Hausdorff dimension of the half-line model above. We do not derive one from the other, yet both are the same drift, as we now show. Let $\tilde \mu \coloneqq \sum_a \tilde w_a \delta_{z_a}$ be the spectral measure of $U$ seen from the last Krylov vector, $\tilde w_a \coloneqq \abs{ \bra{K_{D-1}}\ket{z_a}}^2 = q_{D-1}(z_a).$ Equation~\eqref{em:last} gives it in closed form, 
\begin{equation}
  \tilde w_a = \dfrac{b_a / w_a }{ \sum_{j}b_j/w_j }, \qquad \bar p_{D-1} = \sum_{a = 1}^{D} w_a \tilde w_a, 
\end{equation}
so the last-site occupation is the overlap of the two measures. Since $\mathrm{P}^1 = \mathrm{P}_\blk/ M_{D-1}$ [Eq.~\eqref{eq:Palm_meas}] and $q_{D-1} = M_{D-1}/\sum_m M_m$ at the sampled eigenmode, 
\begin{equation}\label{eq:y_bulk}
  y(D) = D\, \mathrm{E}^1\! \left[ \dfrac{M_{D-1}}{\sum_{m=0}^{D-1} M_m} \right] = D\, \mathrm{E}_\blk\!\left[ \dfrac{1}{\sum_{m=0}^{D-1}M_m} \right]
\end{equation}
at every finite $D$. Weighting by the intensity at the last site thus undoes the Palm tilt; this is the lattice form of the change of measure in Eq.~\eqref{em:tilt}. Under $\mathrm{P}^1$, the ratios read from the last site, $1/X_k$, are independent with law $\mathrm{BetaPrime}(c_k+1, c_k)$, which is just the law of $X_k$ under $\mathrm{P}_\blk$. In this sense, the two laws exchange their roles when the chain is reversed, so the eigenmode sampled with its weight in $\mu$ is an unweighted one for $\tilde \mu$, and vice versa. Equation~\eqref{eq:finite-ladder-reverse} gives the reversal realization by realization.

Under $\mathrm{P}^1$, the intensity at a distance $r$ from the last site, relative to the one at that site, is $M_{D-1-r}/M_{D-1}$, and $\mathrm{E}^1[ \ln(M_{D-1-r}/M_{D-1}) ] = \sum_{j=1}^{r}1 /(cj) \simeq \eta^2 \ln r$. Therefore, the sampled eigenmode typically has a small weight at the last site for every $\beta$, and the mean $y(D)$ is carried by \emph{rare} eigenmodes. The sign of this growth is reversed under the weighting of Eq.~\eqref{eq:y_bulk}, and the continuum limit of the exponent of $Z_s$ in Eq.~\eqref{em:tilt} is $\sqrt{2}\eta B(u) - (1-\eta^2)u$ for $u <s$. Hence, the site count contributes $-1$ and the tilt $+\eta^2$. By Eq.~\eqref{eq:dufre} and Brownian scaling, for $A\coloneqq \int_0^{\infty} e^{a B(u) - bu}\dd u$, $b>0$, $1/A \overset{d}{=} (a^2/2)\gamma_{2b/a^2}$, so $\mathrm{E}[1/A] = b$ for every $a$. Consequently, the noise strength drops out, and the endpoint density equals the drift, $g_\beta(1^-) = 1 - \eta^2$ for $\eta^2 <1$; the same identity at $s = 0$ gives $g_\beta(0^+) = 1 + \eta^2$. The drift $1-\eta^2 = 1 - 2/\beta$ is also the exponent that fixes the dimension of the half-line spectral measure for $\beta>2$~\cite{Simon_2005_2,brow3}. The coincidence of the two values thus rests on the identity $\mathrm{E}[1/A] = b$; that is, on a logarithm of the intensity with independent increments. Whether the endpoint value still equals the spectral dimension when the reflection amplitudes are correlated is open.

\section{The plateau and the Krylov distribution}
\subsection{The plateau coefficient}\label{sm:plateau}
In this section, we derive the scaled mean position~\eqref{eq:F} from which the integer moments of the ensemble-averaged mean Krylov probability density $g_\beta(x)$ can be found. We follow, to a certain extent, the notation of Ref.~\cite{Matsumoto_2005}. From here on, $\mu$ denotes a drift, equal to $1+\beta/2$ in the final formulas, and no longer the spectral measure of Eq.~\eqref{eq:measure}.

Let $B = \{ B(t) : t\geq 0 \}$ be a one-dimensional Brownian motion with $B(0) = 0$, defined on a probability space $(\Omega, \mathcal{F}, P_B)$, and let $B^{(\mu)} =  \{ B_t^{(\mu)} = B(t) + \mu t, t\geq 0 \}$ be the associated Brownian motion with a constant drift $\mu \in \mathbb{R}$. We shall use additive functional $A^{(\mu)} = \{ A_t^{(\mu)} \}$, defined by
\begin{equation}\label{eq:functionalA}
  A_t^{(\mu)} \coloneqq \int_0^t\exp(2B_s^{(\mu)})\dd s ,\quad t \geq 0. 
\end{equation}
The convergence of the profile to Eq.~\eqref{eq:gibbs} is shown in Sec.~\ref{sm:convergence}. Starting from Eq.~\eqref{em:Z}, and using the property $(1/\sqrt{\sigma})B(\sigma x) \overset{d}{=} B(x)$ for $\sigma>0$, we have under $\mathrm{P}^1$, 
\begin{equation}
  \frac{1}{D}\sum_{m=0}^{D-1}M_m \Longrightarrow \mathcal{Z}_\infty \overset{d}{=} \beta A_\infty^{(-\mu)}, \quad \mu = 1 + \dfrac{\beta}{2},
\end{equation}
where the perpetual functional obeys Dufresne's law~\cite{Dufresne01011990}
\begin{equation}\label{eq:dufre}
  A_\infty^{(-\mu)} \overset{d}{=} \dfrac{1}{2\gamma_\mu}.
\end{equation}
Then, in terms of $x$, the rescaled Krylov profiles converge to (see below for a proof)
\begin{equation}
  Dq_n = \dfrac{M_n}{D^{-1}\sum_m M_m} \Longrightarrow \mathcal{G}_\beta(x) = \frac{1}{\mathcal{Z}_\infty}\exp[\frac{2}{\sqrt{\beta}}B(s) - \frac{2}{\beta}s], \quad s = \ln\dfrac{1}{1-x}. 
\end{equation}
The ensemble-averaged mean Krylov probability density is then
\begin{equation}
  g_\beta(x)= \mathrm{E}[\mathcal{G}_\beta(x)],
\end{equation}
where the expectation is taken with respect to the Brownian motion. 

To evaluate the average plateau, we utilize the midpoint coordinate $x_n \coloneqq \frac{n+1/2}{D} \in (0,1).$ Then, from Eq.~\eqref{eq:main1}, the ensemble average of the mean spread is
\begin{equation}
  \mathrm{E}[\bar C_D] = \mathrm{E}^{1}\!\left[\sum_{n=0}^{D-1}n q_n(1) \right] = \mathrm{E}^1\!\left[ \sum_{n=0}^{D-1}x_n Dq_n(1) \right] - \dfrac{1}{2}.
\end{equation}
This identity is exact. The sharper form $\mathrm{E}[\bar C_D] = P(\beta)D - 1/2 + o(1)$ would follow from $\mathrm{E}^1[\sum_n x_n q_n(1)] = P(\beta) + o(1/D)$, which is supported by deterministic evaluation of the finite chain (Sec.~\ref{sm:offset}) but not proved; this sharper form is used for the comparison in Fig.~\ref{fig:A3} of the Letter. To leading order, we calculate
\begin{equation}
  \mathrm{E}[\bar C_D/D] = \mathrm{E}^1\!\left[ \sum_{n=0}^{D-1}x_n q_n(1) \right] - \frac{1}{2D} \xrightarrow[D\rightarrow \infty]{}\mathrm{E}\!\left[\int_0^1x \mathcal{G}_\beta(x)\dd x \right] = \int_0^1xg_\beta(x)\dd x \eqcolon P(\beta).
\end{equation}
The stochastic integrals underlying Eqs.~\eqref{eq:F} and~\eqref{eq:density} are developed in Refs.~\cite{Dufresne01011990,yor,Matsumoto_2005,Comtet_1995,Hariya2004}; their normalization in Eq.~\eqref{em:Z} is needed for the Krylov moments. In the variable $x = 1-e^{-s}$, using $1-x=e^{-s}$ and $\dd x = e^{-s}\dd s$,
\begin{equation}\label{eq:coeff}
  P(\beta) = 1 - \mathrm{E}\!\left[\dfrac{1}{\mathcal{Z}_\infty}\int_0^\infty e^{-s}\exp[\frac{2}{\sqrt{\beta}}B(s) - \left(1+\frac{2}{\beta}\right)s]\dd s \right] \equiv 1- F(\mu,\beta),
\end{equation}
where
\begin{equation}\label{eq:eff}
  F(\mu,\beta) \coloneqq \mathrm{E}\!\left[\dfrac{\int_0^\infty e^{-\beta s}e^{2B_s^{(-\mu)}}\dd s}{A_\infty^{(-\mu)}} \right] = \mathrm{E}\!\left[\dfrac{A_T^{(-\mu)}}{A_\infty^{(-\mu)}} \right], \quad \mu = 1 + \dfrac{\beta}{2}, 
\end{equation}
where we have rescaled $s\to\beta s$ as in Eq.~\eqref{em:Z}, $T\equiv T_\beta \sim \mathrm{Exp}(\beta)$ is independent of $B$, the first $\mathrm{E}[\cdot]$ is with respect to the Brownian motion measure $\mathrm{P}_B$, and the second expectation is with respect to the joint measure $\mathrm{P}_B \otimes \beta e^{-\beta \tau}\dd \tau$. 

Next, split at $T \equiv T_\beta$: $A_\infty^{(-\mu)} = A_{T}^{(-\mu)} + e^{2 B_T^{(-\mu)}}\tilde A$, where $\tilde A$ is an independent copy of $A_\infty^{(-\mu)}$. Using this in Eq.~\eqref{eq:eff} and dividing by $e^{2 B_T^{(-\mu)}}$ yields
\begin{equation}
  \dfrac{A_{T}^{(-\mu)}}{A_\infty^{(-\mu)}} = \dfrac{G}{G + \tilde A}, \quad G \coloneqq e^{-2 B_T^{(-\mu)}}A_T^{(-\mu)} = \int_0^Te^{-2(B_T^{(-\mu)} - B_r^{(-\mu)})}\dd r
\end{equation}
Changing coordinates in the integral of $G$ by $T - r = u$ and since $\exp[2(B(T-u) - B(T) + \mu u)] \overset{d}{=}\exp(2B_u^{(\mu)})$, we get $G\overset{d}{=}A_T^{(\mu)}$. With the independent tail $\tilde A$ of the preceding display, 
\begin{equation}
   \dfrac{A_{T}^{(-\mu)}}{A_\infty^{(-\mu)}} \overset{d}{=} \dfrac{A_T^{(\mu)}}{A_T^{(\mu)} + \tilde A}.
\end{equation}
Let us recall~\cite{Matsumoto_2005}
\begin{equation}
  A_{T_\beta}^{(\mu)} \overset{d}{=} \dfrac{X}{2\gamma_{b-\mu}}, \quad X\sim \mathrm{Beta}(1,b), \quad \gamma_{b-\mu} \sim \mathrm{Gamma}(b-\mu,1), 
\end{equation}
with
\begin{equation}\label{eq:b_c}
  b = \dfrac{1}{2}\left( \sqrt{2\beta + \mu^2} + \mu \right), \quad b-\mu = \dfrac{1}{2}\left( \sqrt{2\beta + \mu^2} - \mu \right),
\end{equation}
Using Dufresne's law for $\tilde A$ and taking $X$, $\gamma_{b-\mu}$, and $\gamma_\mu$ mutually independent gives
\begin{equation}
 \dfrac{A_T^{(\mu)}}{A_T^{(\mu)} + \tilde A} \overset{d}{=} \frac{X R}{1+X R}, \quad R \coloneqq \dfrac{\gamma_{\mu}}{\gamma_{b-\mu}}.
\end{equation}
 Then, from the beta-gamma algebra, $W \coloneqq \gamma_\mu/(\gamma_\mu + \gamma_{b-\mu}) \sim \mathrm{Beta}(\mu,b-\mu)$; so we can write $R \coloneqq W/(1-W)$ and
\begin{equation}
  \frac{XR}{1+XR} = \frac{XW}{1-W(1-X)}.
\end{equation}
Since $W<1$ and $X>0$ almost surely, $W(1-X) \in[0,1)$ almost surely. Hence, we can use the geometric series in the above equation,
\begin{equation}\label{eq:ratios}
  \dfrac{XW}{1-W(1-X)} = \sum_{k\geq 0}X(1-X)^k W^{k+1},
\end{equation}
with 
\begin{equation}
  \mathrm{E}[W^{k+1}] = \dfrac{(\mu)_{k+1}}{(b)_{k+1}}, \quad \mathrm{E}[X(1-X)^k] = \dfrac{b}{(b+k)(b+k+1)}. 
\end{equation}
Replacing these expectations into the expectation of Eq.~\eqref{eq:ratios}, and using $(\mu)_{k+1} = \mu(\mu+1)_k$, $(b)_{k+1} = b(b+1)_{k}$, and $(b+1)_k(b+k+1) = (b+1)(b+2)_k$, gives us
\begin{equation}\label{eq:eff2}
  F(\mu,\beta) = \dfrac{\mu}{b+1}\sum_{k\geq 0}\dfrac{(\mu+ 1)_k}{(b+k)(b+2)_k}.
\end{equation}
To obtain the closed form, write $(b+k)^{-1} = (b)_k/[b(b+1)_k]$ and insert $1 = (1)_k/k!$ to get
\begin{equation}\label{eq:form_F}
  F(\mu,\beta) = \dfrac{\mu}{b(b+1)}\sum_{k\geq 0}\dfrac{(1)_k(b)_k(\mu+1)_k}{(b+1)_k(b+2)_k}\dfrac{1^k}{k!} = \dfrac{\mu}{b(b+1)}{}_3F_2\left(\begin{matrix} 1,b,\mu+1 \\ b+1,b+2 \end{matrix} ; 1\right),
\end{equation}
with $b$ given as in Eq.~\eqref{eq:b_c}. This yields the plateau coefficient in Eq.~\eqref{eq:F}. The same derivation with $T\sim\mathrm{Exp}(q)$ gives $F(\mu,q)$ for any $q>0$, with $\beta$ replaced by $q$ in $b$.

\subsection{The ensemble-average Krylov distribution}
For $\lambda>1$, replacing $e^{-s}$ by $e^{-(\lambda-1)s}$ in the numerator of Eq.~\eqref{eq:coeff} replaces $\mathrm{Exp}(\beta)$ by $\mathrm{Exp}(q)$ with $q=(\lambda-1)\beta$, which gives
\begin{equation}\label{sm:moment-inte1}
  \int_0^1 (1-x)^{\lambda-1}g_\beta(x) \dd x = F(\mu, (\lambda-1)\beta), 
\end{equation}
where the endpoint $\lambda=1$ follows by continuity, with $F(\mu,0)=1$. In Eq.~\eqref{eq:eff2}, $b = (\sqrt{2q +\mu^2}+\mu)/2$ with $q = (\lambda-1)\beta$. Writing
\begin{equation}
  \frac{1}{b + k} = \int_0^1 t^{b+k-1}\dd t, \quad \frac{k!\Gamma(b+1)}{\Gamma(b+k+2)} = \int_0^1 u^k(1-u)^b\dd u,
\end{equation}
and replacing these in Eq.~\eqref{eq:eff2} yields
\begin{equation}
  F(\mu,q) = \mu\int_0^1\int_0^1t^{b-1}(1-u)^b\, \sum_{k\geq 0}\dfrac{(\mu+1)_k(tu)^k}{k!}\dd t \, \dd u. 
\end{equation}
The sum here equals $(1-tu)^{-\mu-1}$ since $tu<1$; hence,
\begin{equation}\label{eq:eff3}
  F(\mu,q) = \mu\int_0^1\int_0^1t^{b-1}(1-u)^b(1-tu)^{-\mu-1}\dd t \, \dd u. 
\end{equation}
Next, let us define a new parameter, rewrite $b$ in terms of it, and introduce a change of variables:
\begin{equation}\label{eq:change}
  \kappa \coloneqq \frac{(\beta-2)^2}{8\beta}, \quad b = \frac{\mu}{2} + \sqrt{\dfrac{\beta}{2}}\sqrt{\lambda+\kappa}, \quad y = \ln\dfrac{1}{t(1-u)} \geq 0. 
\end{equation}
Then, 
\begin{equation}\label{sm:moment-ec2}
  t^{b-1}(1-u)^b = t^{-1}e^{-\mu y/2}e^{-\ell\sqrt{\lambda+\kappa}}, \quad \ell \coloneqq \sqrt{\dfrac{\beta}{2}}y.
\end{equation}
All the $\lambda$-dependence is in $e^{-\ell\sqrt{\lambda+\kappa}}$. Since $\ell>0$ and $\lambda+\kappa >0$, let $\varpi_\ell(s)\coloneqq (\ell/\sqrt{4\pi s^3})e^{-\ell^2/(4s)}$, with Laplace transform
\begin{equation}\label{sm:moment-ec3}
  \int_0^\infty e^{-(\lambda+\kappa)s}\varpi_\ell(s)\dd s = e^{-\ell\sqrt{\lambda+\kappa}}, \quad \lambda > - \kappa. 
\end{equation}
Replacing the right-hand side of this equation into Eq.~\eqref{sm:moment-ec2}, the resulting expression into Eq.~\eqref{eq:eff3}, and swapping the integrals, gives
\begin{equation}\label{eq:inte2}
  F(\mu,q) = \int_0^\infty e^{-\lambda s}\!\left[\mu \int_0^1 \int_0^1t^{-1}e^{-\mu y/2 - \kappa s}\varpi_\ell(s)(1-tu)^{-\mu-1}\dd t \, \dd u \right]\dd s \equiv \int_0^\infty e^{-\lambda s}H(s)\dd s.
\end{equation}
Setting $1-x=e^{-s}$ on the left of Eq.~\eqref{sm:moment-inte1} gives the Laplace transform of $G(s)\coloneqq g_\beta(1-e^{-s})$. Thus,
\begin{equation}
  \int_0^\infty e^{-\lambda s}G(s)\dd s = \int_0^\infty e^{-\lambda s}H(s)\dd s = F(\mu, (\lambda-1)\beta) < \infty, \quad \lambda \geq 1. 
\end{equation}

Uniqueness of the Laplace transform gives $H=G$ almost everywhere. Continuity of both functions will extend the equality to every $s>0$. Let $N(s) \coloneqq \exp(2B(s)/\sqrt{\beta} - 2s/\beta)$. Then, $G(s)=\mathrm E[N(s)/\mathcal Z_\infty]$, with both factors evaluated on the same Brownian path. By Eq.~\eqref{em:tilt}, $G(s)=\mathrm E[Z_s^{-1}]$. Pathwise continuity of $Z_s^{-1}$ and the integrable bound $Z_s^{-1}\leq Z_0^{-1}$ give continuity of $G$ by dominated convergence. Before proving continuity of $H$, we can write it in a more convenient form by replacing $ y$ back as in Eq.~\eqref{eq:change}, and changing to the variable $v = 1-u$. After some algebraic steps, 
\begin{equation}
  H(s) = \frac{\mu}{4}\sqrt{\dfrac{2\beta}{\pi}}e^{-\kappa s}s^{-3/2}\int_0^1\int_0^1 t^{\mu/2-1}v^{\mu/2}(1-t+tv)^{-\mu - 1}\ln\frac{1}{vt}\exp[-\frac{\beta \ln^2(tv)}{8s}]\dd t \, \dd v. 
\end{equation}
Then, changing coordinates to $w = tv$ with fixed $t$, and then to $w = e^{-r}$, upon simplifications, yields
\begin{equation}
  H(s) = \dfrac{\mu}{4}\sqrt{\dfrac{2\beta}{\pi}}e^{-\kappa s}s^{-3/2}\int_0^\infty re^{-\frac{\beta r^2}{8s}}e^{-(\mu/2+1)r}\mathcal{I}_\mu(e^{-r})\dd r, \quad \mathcal{I}_\mu(w) \coloneqq \int_w^1\dfrac{\dd t}{t^2(1-t+w)^{\mu+1}}. 
\end{equation}
 In short, let us write
 \begin{equation}\label{eq:short}
 H(s) = K_\beta e^{-\kappa s}s^{-3/2}\mathcal{J}(s), \quad \mathcal{J}(s) \coloneqq \int_0^\infty r e^{-\beta r^2/(8s)}h_\beta(r)\dd r, \quad h_\beta(r)\coloneqq e^{-(\mu/2+1)r}\mathcal{I}_\mu(e^{-r}), \quad K_\beta \coloneqq \dfrac{\mu}{4}\sqrt{\frac{2\beta}{\pi}}. 
 \end{equation}
 Recall that $r\in [0,\infty)$, and $w = e^{-r}$. For $t\in [w,1]$, $t^{-2} \leq w^{-2}$, and $1+w -t \geq w$; hence, $\mathcal{I}_\mu(w) \leq w^{-\mu-3}$ and $h_\beta(r) \leq e^{(\mu/2 + 2)r}$. Fix $0 < s_0 \leq s \leq s_1$. The integrand $re^{-\beta r^2/(8s)}h_\beta(r)$ is continuous in $s$ for each $r$, and is bounded by $re^{-\beta r^2/(8s_1)+(\mu/2 + 2)r}\in L^1(0,\infty)$. By dominated convergence, $\mathcal{J}$ is continuous on $[s_0,s_1]$; moreover, the prefactor $e^{-\kappa s}s^{-3/2}$ is continuous for $s>0$. Consequently, $H$ is continuous on $(0,\infty)$, and by Lerch's theorem, $G(s) = H(s)$. 

Returning to the $x$-variable, we have found the integral representation of the ensemble-averaged mean Krylov distribution
\begin{equation}
  g_\beta(x) = \dfrac{\mu}{4}\sqrt{\dfrac{2\beta}{\pi}}(1-x)^\kappa\ln^{-3/2}\biggl(\dfrac{1}{1-x}\biggr)\int_0^{\infty}r\exp[\dfrac{\beta r^2}{8\ln(1-x)}]e^{-(\mu/2+1)r}\mathcal{I}_\mu(e^{-r})\dd r,
\end{equation}
with $\kappa$ as in Eq.~\eqref{eq:change}.

\section{Continuum limit, edge behaviors, and first passage}\label{sm:continuum}
This section contains the arguments summarized in Secs.~\ref{em:moments} and~\ref{em:endpoint} of the End Matter: the convergence of the mean profile and its moments, the asymptotics of the Krylov distribution at both ends of the chain, the first-passage identity, and the approach of the last-site occupation to its limit.

\subsection{Convergence of the mean profile and its moments}\label{sm:convergence}
Under $\mathrm P^1$, the central limit theorem of End Matter Sec.~\ref{em:moments} gives $\ln M_n\Rightarrow2B(s)/\sqrt\beta-2s/\beta$ on every bounded interval of $s$. To pass from finite intervals to the full normalization, use $\mathrm E^1[M_n]=1$. The expected contribution of the last $\epsilon D$ sites to $D^{-1}\sum_n M_n$ is at most $\epsilon+D^{-1}$. Thus, the Riemann sums for the total weight and for its integral against a bounded continuous test function converge jointly in distribution. Their ratio has a positive limiting denominator and is bounded by the supremum of the test function. Averaging gives weak convergence of the mean site profile. Its finite-$D$ monotonicity (End Matter Sec.~\ref{em:palm}) then gives pointwise convergence to the continuous $g_\beta(x)=\mathrm E[\mathcal G_\beta(x)]$ for $0<x<1$, by bounding a site density between its averages over adjacent intervals. Equation~\eqref{em:origin} covers $x=0$. The normalization in Eq.~\eqref{em:Z} has an equilibrium reading: its exponent is the negative of the potential for the equilibrium density \emph{per unit $u$}~\cite{yor,BOUCHAUD1990285}.

Setting $\lambda=j+1$ in Eq.~\eqref{sm:moment-inte1} and expanding $x^k=[1-(1-x)]^k$ gives all integer moments,
\begin{equation}\label{sm:mom}
  \int_0^1 x^k g_\beta(x)\,\dd x =\sum_{j=0}^k (-1)^j \binom{k}{j} F(\mu,\beta j).
\end{equation}
The case $k=1$ gives Eq.~\eqref{eq:F} of the Letter, since $F(\mu,0)=1$.

\subsection{Endpoint behaviors of the Krylov distribution}\label{sm:edge}
By setting $u = (1-t+w)/t$ in $\mathcal{I}_\mu$, and then $u = e^v$, we have the more convenient form of $h_\beta$,
\begin{equation}
  h_\beta(r) = \sech^{\beta/2+3}\left(\dfrac{r}{2}\right)\int_0^r \cosh(\dfrac{\beta v}{4})\cosh^{\beta/2+2}\left(\dfrac{v}{2}\right)\dd v,
\end{equation}
which is positive, regular at $r = 0$, and with more transparent asymptotics. 

\noindent
\textbf{Small-$s$ asymptotics.} Let us start with the asymptotics near the seed $x\to 0^+$, that is, $s\to 0^+$. The Gaussian term in $\mathcal{J}(s)$ of Eq.~\eqref{eq:short} localizes the integral around $r = 0$, with $r = O(\sqrt{s})$. Expanding $h_\beta(r)$ around $r = 0$ gives
\begin{equation}
  h_\beta(r) = r + \dfrac{\beta^2 -4\beta -28}{96}r^3 + \dfrac{\beta^4-8\beta^3 - 40\beta^2 +480\beta + 1808}{30720}r^5 + O(r^7). 
\end{equation}
Replacing this into $\mathcal J(s)$ [Eq.~\eqref{eq:short}], integrating term-wise, and then expanding the exponential and algebraic factors appearing in $H(s)$ gives
\begin{equation}
H(s) = \left(1 + \dfrac{2}{\beta} \right)\left[ 1 - \frac{4}{\beta}s + \dfrac{2(\beta+8)}{\beta^2}s^2 + O(s^3) \right], \quad s\to 0^+. 
\end{equation}
Therefore, 
\begin{equation}
\lim_{x\to 0^+}g_\beta(x) = 1 + \dfrac{2}{\beta}.   
\end{equation}

\noindent
\textbf{Large-$s$ asymptotics.} The asymptotic behavior of $h_\beta(r)$ for large $r$ is 
\begin{equation}
  h_\beta(r) \sim \dfrac{1}{1+\beta/2}e^{\delta r}, \quad \delta \coloneqq \dfrac{\beta-2}{4}. 
\end{equation}
To prove $e^{-\delta r}h_\beta(r)\to1/\mu$, put $w=e^{-r}$, so that $e^{-\delta r}h_\beta(r)=w^\mu\mathcal I_\mu(w)$. The substitution $t=1+w-wv$ gives
\begin{equation}
  w^\mu\mathcal I_\mu(w)=\int_1^{1/w}\frac{\dd v}{v^{\mu+1}(1+w-wv)^2}.
\end{equation}
On $1\leq v\leq1/(2w)$, the factor $(1+w-wv)^{-2}$ is at most 4 and tends to one as $w\to0$, so dominated convergence gives $\int_1^\infty v^{-\mu-1}\dd v=1/\mu$ for this part. On $1/(2w)<v\leq1/w$, we have $v^{-\mu-1}\leq(2w)^{\mu+1}$ and $\int(1+w-wv)^{-2}\dd v\leq w^{-2}$, so this part is at most $2^{\mu+1}w^{\mu-1}\to0$ because $\mu>1$. For $0<\beta <2$, $\delta <0$, and $h_\beta(r)$ decays as $r$ grows. Since $\exp[-\beta r^2/(8s)]$ increases to one as $s\to\infty$, monotone convergence gives
\begin{equation}
  \mathcal{J}(s) \sim L_\beta \coloneqq \int_0^\infty r h_\beta(r)\dd r.
\end{equation}
To find this term in an explicit form, define the Mellin transform
\begin{equation}
  M_\mu(p) \coloneqq \int_0^1w^{p-1}\mathcal{I}_\mu(w)\dd w, \quad p>\mu. 
\end{equation}
Since $w = e^{-r}$, we have
\begin{equation}
  L_\beta = -\pdv{M_\mu(p)}{p}\biggl\vert_{p=1+\mu/2}.
\end{equation}
A closed form of $M_\mu(p)$ can be easily found by replacing $(1-t+w)^{-\mu-1} = \sum_{n\geq 0}\frac{(\mu+1)_n}{n!}(t-w)^n$ in $\mathcal{I}_\mu(w)$, and exchanging the integrals, one gets
\begin{equation}
  M_\mu(p) = \dfrac{1}{p(p-1)}{}_3F_2(1,p-1,\mu+1;p,p+1;1). 
\end{equation}
Notice that $M_\mu(p)$ is related to Eq.~\eqref{eq:form_F} via $F(\mu,q) = \mu M_\mu(b+1)$, with $b = \sqrt{\mu^2+2q}/2 + \mu/2$, and $\mu = \beta/2+1$.

The asymptotic behavior of $H(s)$ as $s\to \infty$ is then
\begin{equation}
  H(s) \sim C_\beta s^{-3/2}e^{-\kappa s}, \quad C_\beta \coloneqq -\dfrac{\mu}{4}\sqrt{\dfrac{2\beta}{\pi}}M_\mu'(1 + \mu/2), \quad s \to \infty, \quad 0 < \beta < 2.
\end{equation}
In terms of $x$, we get the asymptotics of $g_\beta(x)$ as $x \to 1^-$,
\begin{equation}
  g_\beta(x)\sim C_\beta(1-x)^\kappa \ln^{-3/2}\left(\dfrac{1}{1-x} \right), \quad x \to 1^-, \quad 0 < \beta < 2.
\end{equation}

For $\beta = 2$, $\delta = 0$, and so $h_\beta(r) \to 1/2$ as $r\to\infty$. 
Substituting $r=2\sqrt{s}\rho$ in $\mathcal{J}(s)$ and using dominated convergence, we get
\begin{equation}
  H(s)\sim \dfrac{1}{\sqrt{\pi s}}, \quad s\to \infty, \quad \beta = 2, 
\end{equation}
and 
\begin{equation}
  g_\beta(x) \sim \dfrac{1}{\sqrt{\pi}}\ln^{-1/2}\left( \dfrac{1}{1-x}\right). 
\end{equation}

Finally, for $\beta >2$, $\delta >0$ and $h_\beta(r)$ grows exponentially, so the exponential in $\mathcal{J}(s)$ cannot be Taylor-expanded under the integral. Instead, the integrand contains an exponential of $-\beta r^2/(8s)+\delta r$, where the dominant contribution comes from the saddle $r_* = (\beta-2)s/\beta$, which lies in the integration domain and tends to infinity as $s\to \infty$. Then, Laplace's method gives
\begin{equation}
  H(s) \sim \dfrac{\beta-2}{\beta}, \quad s \to \infty, \quad \beta > 2, 
\end{equation}
and the same for $g_\beta(x)$ as $x\to 1^-$. Equation~\eqref{em:tilt} of the End Matter gives an independent proof of this limit.

\subsection{First-passage identity}\label{sm:passage}
To prove Eq.~\eqref{em:maximum}, set $t=s/\beta$ and $A_t^{(\pm \mu)}$ as in Eq.~\eqref{eq:functionalA} and recall Eq.~\eqref{eq:dufre}. Split this integral at $t$, divide both pieces by the integrand at $t$, and reverse the Brownian increments on $[0,t]$, as in Sec.~\ref{sm:plateau}. The remaining tail is independent of the reversed interval. Thus, the left side of Eq.~\eqref{em:maximum} equals
\begin{equation}
  \mathrm E\!\left[\frac{\widetilde A_\infty^{(-\mu)}}{\widetilde A_\infty^{(-\mu)}+A_t^{(\mu)}}\right]
\end{equation}
with independent integrals. For a diffusion in $W(u)+hu$, its probability, in a fixed environment, to hit $y>0$ is $I_-/(I_-+I_y)$, where $I_-=\int_{-\infty}^0 e^{W(u)+hu}\dd u$ and $I_y=\int_0^y e^{W(u)+hu}\dd u$~\cite{KawazuTanakaMaximum1993}. Setting $u=4v$, $y=4t$, and $h=\mu/2$ matches the two pairs of integrals in law.

Kawazu and Tanaka derived the three tails of the all-time maximum in 1993~\cite{KawazuTanakaMaximum1993}. The three laws, restated by Matsumoto and Yor in Theorem~4.1 of their survey~\cite{MatsumotoYor2005II}, give an independent check of Eq.~\eqref{em:continuumedge}. For $h>1$, $\Pr(\xi_h>y)\simeq[2(h-1)/(2h-1)]e^{-(h-1/2)y}$, and for $h=1$, $\Pr(\xi_h>y)\simeq(2/\pi)^{1/2}y^{-1/2}e^{-y/2}$. With $y=4s/\beta$, these become $(1-2/\beta)e^{-s}$ and $(\pi s)^{-1/2}e^{-s}$, the leading behaviors of the left side of Eq.~\eqref{em:maximum} for $\beta>2$ and $\beta=2$. For $h<1$, the tail is proportional to $y^{-3/2}e^{-h^2y/2}$, which gives $s^{-3/2}e^{-(1+\kappa)s}$ for the cumulative occupation.

To recover the density, write the cumulative occupation as $Q(s)=\int_s^\infty e^{-u}G(u)\,\dd u$, with $G(u)=g_\beta(1-e^{-u})$ decreasing. The averages of $G$ with weight $e^{-u}$ over $[s,s+d]$ and $[s-d,s]$ bound $G(s)$ from below and above, respectively. If $Q(s)\sim A e^{-\lambda s}s^{-\nu}$, taking $s\to\infty$ and then $d\to 0^+$ in these bounds gives $G(s)\sim\lambda e^s Q(s)$.

The later work on passage \emph{times} in a drifted Brownian environment studies a different observable~\cite{Kawazu2002}.

\subsection{The last site on the exact chain}\label{sm:lattice}
The limiting-value argument of End Matter Sec.~\ref{em:endpoint} does not fix the rates at which the discrete endpoint approaches its limit. With $s=\ln D$, Eq.~\eqref{em:continuumedge} suggests a dependence $D^{-\kappa}(\ln D)^{-3/2}$ for $\beta<2$ and $1/\sqrt{\pi\ln D}$ for $\beta=2$. At finite $D$, a recursion for the distribution of one positive variable gives the exact endpoint. Set $S_k=\sum_{j=0}^k M_j/M_k$. By Eqs.~\eqref{eq:main1} and~\eqref{em:ratios},
$$
 S_0=1,\qquad S_{k+1}=1+\frac{S_k}{X_k},\qquad
 y(D)=D\,\mathrm E^1\!\left[\frac{1}{S_{D-1}}\right],
$$
where $X_k$ is independent of $S_k$ and has the beta-prime law in Eq.~\eqref{em:ratios}. Deterministic density transfer on the exact chain extends to $D=2^{20}$ [Fig.~\ref{fig:A2} of the Letter]. For $\beta=3,4,6,8$, extrapolated endpoint values are $0.3368(78)$, $0.5003(16)$, $0.6667(3)$, and $0.7500(1)$, respectively. At $\beta=2$, the fit in Fig.~\ref{fig:A2}(b) uses $a+b/\ln D+c/\ln^2 D$ for $\sqrt{\pi\ln D}\,y(D)$ at $D\geq2^{14}$. The fitted coefficient of $1/\sqrt{\pi\ln D}$ is $1.004(15)$, with $\sqrt{\pi\ln D}\,y(D)$ still near $1.05$ at the largest size. For $\beta=1/2,1,4/3$, the fitted powers are $0.562(3)$, $0.120(7)$, and $0.030(11)$, against continuum exponents $0.5625$, $0.1250$, and $0.0417$. Parentheses here give the spreads across fit variants.

\paragraph{Methods.} The computation uses the form of Eq.~\eqref{eq:y_bulk}. Under $\mathrm{P}_\blk$, the ratios are independent with $X_k \sim \mathrm{BetaPrime}(c_k+1,c_k)$, and $y(D) = D\, \mathrm{E}_\blk[1/Z_0]$, where $Z_j = \sum_{n \geq j}M_n / M_j$ obeys $Z_{D-1} = 1$ and $Z_j = 1 + X_j Z_{j+1}$. With $v_j = \ln(Z_j-1)$, one step reads $v_j = \ln X_j + \ln (1 + e^{v_{j+1}})$, where $X_j$ is independent of $v_{j+1}$. We propagate the probability density of $v_j$ from the last link to the first on a uniform grid with step $0.04$. The density of $\ln(1 + e^{v_{j+1}})$ follows by a change of variable. The convolution with the density of $\ln X_j$ is evaluated in Fourier space, with the exact characteristic function $\mathrm{E}[X_{j}^{-i\omega}] = \Gamma(c_j + 1 - i\omega)\Gamma(c_j + i\omega)/[\Gamma(c_j+1)\Gamma(c_j)]$, where this density is narrow $(c_j \geq 12)$, and by direct quadrature for the links closer to the end. No random numbers are used. The values cover $D = 2^6, 2^7, \ldots, 2^{20}$ for the eight values of $\beta$ of Fig.~\ref{fig:A1.0} ($D \leq 2^{19}$ at $\beta = 1/2$), 119 values in all. The total probability is conserved to $6.4\times 10^{-10}$. Halving the grid step changes $y(D)$ by at most $1.5\times 10^{-9}$ in relative terms at the 21 pairs $(\beta,D)$ tested ($D = 2^{12}$ and $2^{16}$ for all $\beta$, and $D = 2^{20}$ for $\beta \geq 2$). The sampled chain of Sec.~\ref{sm:sampled} agrees with these values at the 48 common points ($D = 2^{6},\ldots, 2^{11}$): the largest deviation is $2.2$ standard errors, and the sum of the squared deviations in these units is $34.9$.

\paragraph{Fits.} The lattice values carry no statistical errors. The uncertainty quoted for a fitted parameter is therefore the largest deviation from the reference fit ($D \geq 2^{14}$, with the subleading term) among the six fits with $D\geq 2^{12},2^{14}$ or $2^{16}$, with and without the subleading term. For $\beta >2$, we fit $y = y_\infty + \mathsf{a} D^{-\kappa}(\ln D)^{-3/2}(1 + \mathsf d /\ln D)$, with $\kappa = (\beta - 2)^2/(8\beta)$ fixed. For $\beta = 2$, we fit $\sqrt{\pi \ln D}y = a + b/\ln D + c/\ln^2D$, where $c$ is the subleading term. For $\beta <2$, we fit $y = \mathsf a D^{-k}(\ln D)^{-3/2}(1 + \mathsf d_1/\ln D + \mathsf d_2 / \ln^2 D)$, with $k$ free and $\mathsf d_2$ the subleading term. Table~\ref{tab:fits} lists the results. The amplitudes $\mathsf a$ are strongly correlated with the other parameters and are not quoted. The extrapolated limits for $\beta >2$ depend on the assumed form of the approach: a series in $1/\ln D$ in place of the power form gives $y_\infty = 0.316$, $0.496$, $0.670$, and $0.752$ for $\beta = 3, 4,6,8$, so that the value at $\beta = 3$ then lies $5\%$ below $1/3$. The proof of Eq.~\eqref{eq:end} does not rely on these fits.

\begin{table}[htb]
\setlength{\tabcolsep}{7pt}
\caption{Fits to the exact last-site occupation: exponent $k$ ($\beta<2$, continuum value $\kappa$), intercept $a$ ($\beta=2$, continuum value $1$), and limit $y_\infty$ ($\beta>2$, exact value $1-2/\beta$). Spread: largest deviation from the reference fit among the fit variants.}\label{tab:fits}
\begin{tabular}{@{}clccc@{}}
\toprule
$\beta$ & parameter & value & spread & prediction\\
\midrule
$1/2$ & $k$ & $0.5618$ & $0.0027$ & $0.5625$ \\
$1$ & $k$ & $0.1200$ & $0.0075$ & $0.1250$ \\
$4/3$ & $k$ & $0.0303$ & $0.0110$ & $0.0417$ \\
$2$ & $a$ & $1.004$ & $0.015$ & $1$ \\
$3$ & $y_\infty$ & $0.3368$ & $0.0078$ & $0.3333$ \\
$4$ & $y_\infty$ & $0.5003$ & $0.0016$ & $0.5000$ \\
$6$ & $y_\infty$ & $0.6667$ & $0.0003$ & $0.6667$ \\
$8$ & $y_\infty$ & $0.7500$ & $0.0001$ & $0.7500$ \\
\bottomrule
\end{tabular}
\end{table}

\section{Models and numerical tests}\label{sm:models}
\subsection{Kicked tops}
We considered the three models of kicked tops used in Ref.~\cite{kicked1} to study long-time quantum Fisher information and multipartite entanglement, with total \emph{half-integer} spin $j$, $N = 2j+1$ levels, and angular-momentum operators $J_x, J_y, J_z$. The Hilbert space is $\mathcal{H} = \mathbb C^N$, spanned by $\ket{j,m}\equiv \ket{m},$ $m = -j, \ldots, j$. The models we treated in the Letter are thus
\begin{align}
  U_{\text{COE}} &= \exp(-i\dfrac{\lambda}{N}J_z^2)\exp(-iAJ_x), &\quad &\lambda \in [8,12], \quad A\in [1.6, 1.8], \label{eq:Ucoe}\\
  U_{\text{CUE}} &= \exp(-i\dfrac{\lambda'}{N}J_y^2)\exp(-i\dfrac{\lambda}{N}J_z^2)\exp(-i pJ_x), &\quad &\lambda = 10, \quad \lambda' \in [4,7], \quad p \in [1.6,1.8], \label{eq:Ucue}\\ 
  U_{\text{CSE}} &= \exp(-iV)\exp(-iH_0), &\quad & \lambda_0\in [2.35,2.65], \quad \lambda_1 \in [2,3], \quad \lambda_2\in [4,6], \quad \lambda_3 \in [6,9]. 
\label{eq:Ucse}
\end{align}
where
\begin{equation}
  H_0 = 2\dfrac{\lambda_0}{N}J_z^2, \quad V = \dfrac{8\lambda_1}{N^3}J_z^4 + \dfrac{2\lambda_2}{N}\{ J_x,J_z \} + \dfrac{2\lambda_3}{N}\{ J_x, J_y \}. 
\end{equation}

\noindent
\textbf{Couplings and operators.} An \emph{operator} is one set of couplings. For each coupling, $200$ values are drawn independently and uniformly from its range in Eqs.~\eqref{eq:Ucoe}--\eqref{eq:Ucse}. For $U_\mathrm{COE}$ and $U_\mathrm{CSE}$, the 200 draws of each coupling are sorted, and operator $m$ takes the $m$-th smallest value of every coupling, so the couplings of one operator increase together. For $U_\mathrm{CUE}$, $\lambda'$ is assigned in the same way and $p$ is drawn independently of it. The same 200 sets of couplings are used across the four sizes $D = 100, 200, 400, 800$ ($ j = D-1/2$), with initial states drawn independently at each size. 
\noindent

\textbf{Unitary symmetries.} The COE and CUE operators have the unitary symmetry $R_x(\pi) = \exp(-i\pi J_x)$, so each one has two parity sectors of dimension $N/2$, which are taken to be the Krylov dimensions for each. A basis for each sector is $\ket{m}_\pm = (\ket{m} \pm \ket{-m})/\sqrt{2}$ with $m>0$. For the parameters in Eq.~\eqref{eq:Ucse}, the CSE top has no spatial symmetry that further splits the spin space. 

\noindent
\textbf{Antiunitary symmetries.} $U_\text{COE}$ satisfies $T_1U_\text{COE}T_1^{-1}=U_\text{COE}^\dagger$ for the antiunitary operator $T_1 = e^{i AJ_x}K$, where $K$ denotes complex conjugation, and $T_1^2 = 1$ for all $j$. Thus, we treat it in its symmetric frame, implemented by the unitary operator $W = e^{-iAJ_x/2}$, such that $U_s = W U_{\text{COE}}W^\dagger = e^{-iAJ_x/2}e^{-i(\lambda/N) J_z^2} e^{-iAJ_x/2} = U_s^\top$. For the parameters in Eq.~\eqref{eq:Ucue}, the drive has no time-reversal symmetry within either parity sector. Finally, $U_\text{CSE}$ satisfies $T_3U_\text{CSE}T_3^{-1}=U_\text{CSE}^\dagger$ for $T_3 = e^{-iV}e^{-i\pi J_y}K$, $T_3^2 = (-1)^{2j} = -1$ for half-integer $j$. Each eigenvalue of $U_\text{CSE}$ is doubly degenerate, which is the Kramers degeneracy; then, it has $N/2$ distinct eigenphases, each with a two-dimensional eigenspace $\{\ket{z_a}, T_3\ket{z_a} \}$. Therefore, a seed with nonzero projection on every doublet has $N/2$ spectral atoms with the \emph{doublet weights}
\begin{equation}\label{eq:kramers}
  w_a = \abs{\bra{z_a}\ket{\psi_0}}^2 + \abs{\bra{T_3 z_a}\ket{\psi_0}}^2,
\end{equation}
and the Krylov chain has $D = N/2$ sites.

\noindent
\textbf{Initial states.} For $U_\mathrm{COE}$, the initial states are real Gaussian vectors in one parity sector of the symmetric frame, $\ket{\psi_0}\propto \sum_{m>0} g_m\ket{m}_\pm$ with independent standard normal $g_m$; the eigenvectors of $U_s = U_s^\top$ can be chosen real in that frame. For $U_\mathrm{CUE}$, they are complex Gaussian vectors in one parity sector. For $U_\mathrm{CSE}$, they are complex Gaussian vectors in the full space $\mathbb C^N$, with the doublet weights~\eqref{eq:kramers}. Since an orthogonal (unitary) change of basis maps a real (complex) Gaussian vector to a vector with the same law, the weights of these states follow Eq.~\eqref{eq:dir}, with $\beta = 1,2,4$, for every operator, and are independent of its quasienergies. The comparison with the circular ensembles therefore tests the quasienergies of the models and not their eigenvectors. Each parity sector of each operator is one \emph{block}; $U_\mathrm{CSE}$ has one block per operator.

\subsection{Brickwork Floquet circuits}
The brickwork Floquet circuits (henceforth, circuits) are comprised of $L$ qubits on an open chain, $\mathcal{H} = (\mathbb{C}^{2})^{\otimes L}$. A layer $B$ is a tensor product of independent Haar-random $U(4)$ gates, one on each bond of the given parity: even bonds $(1,2), (3,4),\ldots$ or odd bonds $(2,3), (4,5),\ldots$; consecutive layers alternate parity. The gates are drawn \emph{once} and fixed, so repeating the same $V$ below is what makes it a Floquet operator rather than a random circuit~\cite{C1}. One \emph{brick} is the pair of layers $B = B_\text{odd}B_\text{even}$, and $V$ is four bricks, eight gate layers,
\begin{equation}
  V \coloneqq B_4 B_3 B_2 B_1, \quad B_i = B_{\text{odd},i}B_{\text{even},i}.
\end{equation}
We use four bricks per period; at this depth, the ratios of adjacent level spacings match those of circular ensembles at all sizes studied (Fig.~\ref{fig:rbar}). For the three classes, we use the following construction. For any antiunitary $\Theta$ with $\Theta^2 = \pm 1$, and unitary $V$, we set $U = \Theta V^\dagger \Theta^{-1}V$. Then, 
\begin{equation}
  \Theta U \Theta^{-1} = V^\dagger \Theta V \Theta^{-1} = U^\dagger. 
\end{equation}
For the CUE circuit, we used $U_\text{CUE}=V$, with no symmetry imposed on the gates. Its Krylov dimension is $D = 2^L$. The COE circuit we used is 
\begin{equation}
  U_\text{COE} = V^\top V = U_\text{COE}^\top,
\end{equation}
so $\Theta = K$, $KUK^{-1} = U^* = U^\dagger$, with $K^2 = 1$. The Krylov dimension is $D = 2^L$. Lastly, for the CSE circuit, 
\begin{equation}
  U_\text{CSE} = \Theta V^\dagger \Theta^{-1}V, \quad \Theta =\left(\prod_{k=1}^Li Y_k \right)K, \quad L \ \text{odd}, \quad \Theta^2 = (-1)^L = -1.
\end{equation}
Due to the Kramers degeneracy, the Krylov dimension is $D = 2^{L-1}$.

One period of the orthogonal and symplectic circuits thus contains $16$ gate layers: the eight layers of $V$ and their images under $\Theta$, and one period of the unitary circuit contains eight. An operator is one draw of all the gates of $V$; the gates are drawn independently for every operator and every $L$, and each operator is one block. The initial states are real Gaussian vectors in the computational basis for $U_\mathrm{COE}$ and complex Gaussian vectors for $U_\mathrm{CUE}$ and $U_\mathrm{CSE}$, with doublet weights as in Eq.~\eqref{eq:kramers} for the latter. As for the tops, their weights follow Eq.~\eqref{eq:dir}.

\subsection{Sizes, controls, and the regular top}\label{sm:numerics}
The sizes are $D=100,200,400,800$ for each top, $D=2^6,\ldots,2^{11}$ for orthogonal and unitary circuits, and $D=2^6,2^8,2^{10}$ for symplectic circuits ($L = 6, \ldots, 11$ and $L = 7, 9, 11$ qubits). For the circular-ensemble data of Fig.~\ref{fig:A1.0}, the Krylov basis for each choice of phases and weights is that of its spectral measure. The bin comparison in Fig.~\ref{fig:A4}(a) and the single-site endpoint test in Fig.~\ref{fig:A2}(a) resolve different length scales.

The independent-phase control uses 400 spectra and five weight vectors per spectrum and size, with $64\leq D\leq2048$ (the ten sizes of Sec.~\ref{sm:sampled}). Fitting the spread divided by $D$ to $P_{\mathrm P}+b_{\mathrm P}/D$ gives $P_{\mathrm P}=0.3232(1),0.3939(1),0.4412(1)$ and $b_{\mathrm P}=-0.62(3),-0.60(2),-0.55(2)$ for $\beta=1,2,4$ (weighted least squares over the ten sizes; Sec.~\ref{sm:offset}). Over the sizes examined, the end retains roughly $26$--$73\%$ of the uniform share across these classes. The inverse-variance means over the ten sizes are $y = 0.257(3)$, $0.521(3)$, and $0.730(3)$ for the weights $\beta = 1,2,4$, with no detectable dependence on $D$ ($\chi^2 = 10.6, 8.8$ and $6.0$ on 9 degrees of freedom). For the weights of $\beta = 4$, End Matter Eq.~\eqref{em:last} and Jensen's inequality give $y(D) \geq D/(2D-1) >1/2$ for any phases independent of the weights, so a finite last-site occupation is then guaranteed. Figure~\ref{fig:A4}(c) shows a redistribution over a finite part of the chain. Changing an individual site's occupation by $O(1/D)$ could change the scaled mean spread by only $O(1/D)$. A closed formula for $P_{\mathrm P}$ remains open.

For the regular top, $U=\exp[-ik_xJ_x^2/(2J)]\exp[-iA J_x]$ with integer $J$, $A=1.7$, and $k_x$ uniform on $[10,12]$. Both factors are functions of $J_x$, so $[U,J_x] = 0$: the eigenvectors are the eigenstates $\ket{m}_x$ of $J_x$, and the eigenphases are $\theta_m = -k_x m^2/(2J) - Am$ modulo $2\pi$. We used $J = 25, 50,100,200, 400,800$, with 200 realizations of $k_x$ per size and four real and four complex Gaussian initial states each. The real and complex Gaussian initial states used here occupy the full $D_K=2J+1$ multiplet. Their scaled plateaus follow the corresponding independent-phase fits within $1.1\times10^{-3}$ (complex states: at most $8.1\times 10^{-4}$ over the six sizes, with a sum of squared residuals of $4.9$ in units of the standard errors; real states: $1.1\times 10^{-3}$ and $7.2$). The spin-coherent comparison of Ref.~\cite{streamed} uses a reflection sector as well as a different initial state. A spin-coherent state has binomial weights on the $\ket{m}_x$, concentrated on of order $\sqrt{J}$ of them.

Figure~\ref{fig:rbar} shows the mean ratio $\bar r$ of adjacent quasienergy spacings. For each operator, $\bar r$ is first averaged over its symmetry sectors; the standard error $\sigma$ is taken over the 200 operators. At all 27 points, $\bar r$ lies within two standard errors of the large-$D$ circular-ensemble value $r_\beta$~\cite{Atas}, with largest deviation $0.0029$. The regular top follows the value $2\ln 2 -1$ for independent phases [$\bar r = 0.397(4)$ at $J = 25$ and $0.384$--$0.388$ at the five larger sizes]. 

\begin{figure}[t!]
 \centering
 \includegraphics[width=\textwidth]{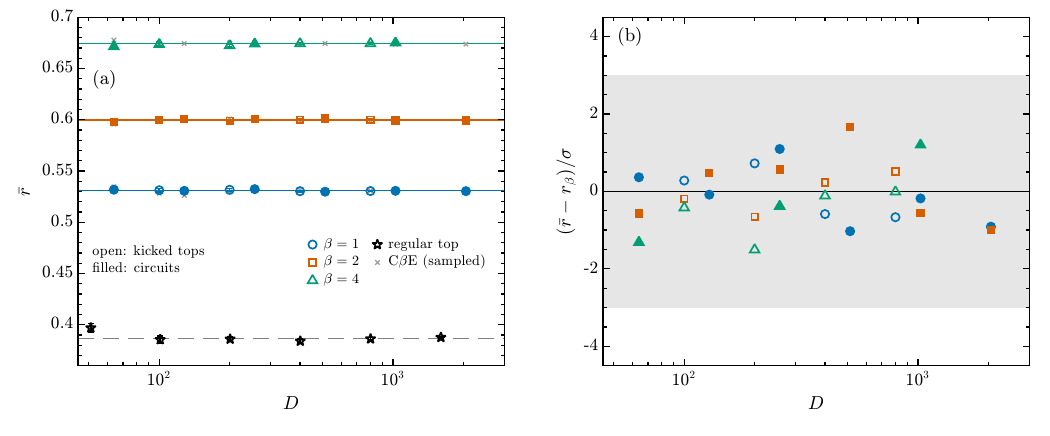}
 \caption{Mean ratio $\bar r$ of adjacent quasienergy spacings. (a) Kicked tops (open symbols), brickwork circuits (filled symbols), and the regular kicked top against the Krylov dimension $D$ ($D_K = 2J+1$ for the regular top), with one-standard-error bars over 200 operators. Solid lines: large-$D$ circular-ensemble values $r_\beta = 0.5307$, $0.5996$, and $0.6744$ for $\beta = 1,2,4$~\cite{Atas}. Dashed line: independent phases, $2\ln 2 -1$~\cite{Atas}. Crosses: sampled C$\beta$E at the same $D$. (b) Deviations $(\bar r - r_\beta)/\sigma$ of the 27 chaotic points from the solid lines, where $\sigma$ is the standard error of $\bar r$. The shaded band marks three standard errors; all points lie within $1.66$.  }
 \label{fig:rbar}
\end{figure}

\subsection{Samples and random numbers}\label{sm:samples}
Table~\ref{tab:samples} lists the samples behind every data point of the Letter. For the models, a point is one model at one size. The circular-ensemble samples of Sec.~\ref{sm:sampled} and the independent-phase control use the ten sizes of the two reference ensembles are $D = 64$, 100, 128, 200, 256, 400, 512, 800, 1024, 2048.

\begin{table}[htb]
\setlength{\tabcolsep}{7pt}
\caption{Samples. A block is one symmetry sector of one operator. For the last two rows, the entries are spectra and their corresponding weight vectors.}\label{tab:samples}
\begin{tabular}{@{}llccc@{}}
\toprule
family & sizes & blocks per operator & initial states per block & operators\\
\midrule
kicked top, $\beta=1,2$ & $D=100,200$ & 2 & 400 & 200\\
 & $D=400,800$ & 2 & 200 & 200\\
kicked top, $\beta=4$ & $D=100,200$ & 1 & 400 & 200\\
 & $D=400,800$ & 1 & 200 & 200\\
circuit, $\beta=1,2$ & $D=2^6,\ldots,2^{11}$ & 1 & 100 & 200\\
circuit, $\beta=4$ & $D=2^6,2^8,2^{10}$ & 1 & 200 & 200\\
regular top & $D_K=51,101,201,401,801,1601$ & 1 & 4 real and 4 complex & 200\\
\midrule
circular ensemble, eight $\beta$ & ten sizes & -- & 50 weight vectors & 200 spectra\\
independent phases, eight $\beta$ & ten sizes & -- & 5 weight vectors & 400 spectra\\
\bottomrule
\end{tabular}
\end{table}

\subsection{Computation of the Krylov occupations}\label{sm:algorithm}
\paragraph{Diagonalization.} Every block is diagonalized by a complex Schur decomposition, whose Schur vectors form an orthonormal eigenbasis also when two eigenvalues are close. Over all blocks of the six chaotic models, $\norm{U^\dagger U-1}_F/\sqrt N\leq8.4\times10^{-15}$ for the $N\times N$ Floquet matrix, the strictly upper-triangular part of the Schur form has Frobenius norm at most $8\times10^{-13}$, and the smallest nearest-neighbor spacing is $1.2\times10^{-3}$ of the mean spacing. For the symplectic models, the eigenphases are sorted and grouped into consecutive pairs. The largest splitting within a pair is $2\times10^{-13}$, and the smallest distance between neighboring pairs is $0.05$ of their mean spacing. The symmetry relations of Sec.~\ref{sm:models} hold to $\norm{U_s-U_s^\top}_F/\norm{U_s}_F\leq10^{-15}$ for the orthogonal top and $\norm{T_3U_\text{CSE}T_3^{-1}-U_\text{CSE}^\dagger}_F/\norm{U_\text{CSE}}_F\leq1.2\times10^{-12}$ for the symplectic top; the corresponding relations for the circuits hold to $10^{-15}$.

\paragraph{Recursion in the spectral representation.} Given the distinct eigenphases $z_a$ and the weights $w_a$ of one initial state, the occupations follow from the values $\phi_n(z_a)$ and $\tilde\phi_n(z_a)$ on the $D$ atoms. Starting from $\phi_0=\tilde\phi_0=1$, orthogonality of $\phi_{n+1}$ to the constant function gives
\begin{equation}\label{eq:alpha_spec}
  \alpha_n^*=\frac{\sum_a w_a z_a\phi_n(z_a)}{\sum_a w_a\tilde\phi_n(z_a)},
\end{equation}
after which Eq.~\eqref{eq:recurrences} and its reversed form advance both polynomials, with $\rho_n$ fixed by $\sum_a w_a\abs{\phi_{n+1}(z_a)}^2=1$. Then $q_n(z_a)=w_a\abs{\phi_n(z_a)}^2$, $\bar p_n=\sum_a w_aq_n(z_a)$, $\bar C_D=\sum_n n\bar p_n$, and the last-site occupation is $D\bar p_{D-1}$. The cost is $O(D^2)$ per initial state.

\paragraph{Accuracy check and backward-stable route.} The recursion is forward unstable when the weights span many orders of magnitude or the phases cluster. We monitor it through the Christoffel identity $w_a\sum_n\abs{\phi_n(z_a)}^2=1$, with the indicator
\begin{equation}\label{eq:eacc}
  \varepsilon=\frac{\sum_a w_a^2\,s_a\,\abs{w_at_a-1}}{\sum_a w_a^2\,s_a},\qquad
  t_a=\sum_{n}\abs{\phi_n(z_a)}^2,\quad s_a=\sum_{n}n\abs{\phi_n(z_a)}^2 .
\end{equation}
An initial state with $\varepsilon>10^{-6}$ ($10^{-5}$ for the circuits) is recomputed as follows. Let $\mathsf H$ be the real Householder reflector that maps the first basis vector to $(\sqrt{w_a})_a$. The matrix $\mathsf H\,\diag(z_a)\,\mathsf H$ is reduced to upper Hessenberg form by a unitary similarity that fixes the first basis vector. The columns of the accumulated transformation are the Krylov vectors in the eigenbasis, so their squared moduli are $q_n(z_a)$. This route is backward stable and costs $O(D^3)$ per state. For the models, this concerns only the orthogonal top ($0$, $3$, $147$, and $1102$ states at $D=100,200,400,800$, out of $1.6\times10^5$, $1.6\times10^5$, $8\times10^4$, and $8\times10^4$) and the orthogonal circuit ($170$ and $650$ of $2\times10^4$ at $D=1024$ and $2048$, at most $3$ at the smaller sizes); for the unitary and symplectic models, there are none. In the sampled circular ensemble, it concerns $97$ to $9972$ of the $10^4$ states per size at $\beta=1/2$ as $D$ grows from $64$ to $2048$, at most $1197$ at $\beta=1$, at most $5$ at $\beta=4/3$, and none for $\beta\geq2$. For independent phases, it is used for $92$--$100\%$ of the states, and for the regular top for most of them. A state that passes the check has a relative error below the tolerance, which is negligible compared with the statistical errors of Secs.~\ref{sm:plateautests} and~\ref{sm:farendtests}.

\subsection{Sampled circular ensemble}\label{sm:sampled}
For each of the eight values of $\beta$ and the ten sizes, we draw 200 sets of reflection amplitudes from Eq.~\eqref{eq:verb}, build the matrix of Eq.~\eqref{eq:CMV}, and diagonalize it. Each spectrum is combined with 50 weight vectors drawn from Eq.~\eqref{eq:dir} independently of it. By the theorem of Killip and Nenciu, these have the joint law of the weights of the first basis vector of the matrix. The occupation of a matrix is the mean over its 50 weight vectors, the estimate is the mean of the 200 matrix values, and the standard error is their standard deviation divided by $\sqrt{200}$. The weight vectors of one matrix share its spectrum and are not treated as independent. All sites of a profile come from the same matrices, and every profile is normalized, so the errors at different sites are correlated. The symbols of Fig.~\ref{fig:A1.0}(a,b) are 15 sites per $\beta$, uniformly spaced in the lattice clock $s_n$, at $D = 2048$. Of the 120 residuals in Fig.~\ref{fig:A1.0}(b), six exceed three standard errors, all within the last five sites of the chain; the largest is $-1.2\times 10^{-2}$ at $\beta = 2$, $n = 2047$. 

We performed three checks of these data. (i) The mean spread follows $P(\beta) - 1/(2D)$ at the 80 pairs $(\beta, D)$: the largest deviation is $2.85$ standard errors, and the sum of the squared deviations in these units is $75.1$. (ii) The spread computed from the first basis vector of each matrix agrees with the mean over its 50 independent weight vectors: the largest difference is $2.7$ standard errors, and the sum of squares is $78.7$ on 80. Only the first basis vector involves the eigenvectors of the matrix, so this tests its construction. (iii) The return probability follows Eq.~\eqref{em:origin}: the sum of squares is $73.4$ on 80, with one deviation of $3.3$ standard errors. At $\beta = 1/2$, most states at large $D$ are computed by the backward-stable route (Sec.~\ref{sm:algorithm}), and the last few sites of that curve in Fig.~\ref{fig:A1.0}(a) are the least accurate in the figure; no statement about the far end at $\beta = 1/2$ rests on the sampled chain. 

\subsection{The offset $-1/2$}\label{sm:offset}
Two computations support our result $\mathrm{E}[\bar C_D] = P(\beta)D - 1/2 + o(1)$. The first is deterministic. The method of Sec.~\ref{sm:lattice}, applied under $\mathrm{P}^1$ and propagating the first moment $\sum_n n q_n$ together with the density, gives $\mathrm{E}[\bar C_D]$ without sampling. With $P(\beta)$ as in Eq.~\eqref{eq:F}, 
\begin{equation}\label{eq:offset_det}
  \mathrm{E}[\bar C_D] - P(\beta)D + \dfrac{1}{2} = \begin{cases}
    1.46\times 10^{-4}, & \beta = 1/2, \ D = 6400, \\
    3.7\times 10^{-5}, & \beta = 1/2, \ D = 25600, \\
    9.4\times 10^{-6}, & \beta = 2, \ D = 25600, \\
    1.8 \times 10^{-6}, & \beta = 8, \ D = 25600. 
  \end{cases}
\end{equation}
These values scale as $1/D$, with coefficients $0.94$, $0.24$ and $0.05$ for $\beta = 1/2,2,8$. Halving the grid step from $0.02$ to $0.01$ changes $\mathrm{E}[\bar C_D]$ by $5\times 10^{-10}$ at $\beta = 1/2$, $D = 6400$, by $6\times 10^{-9}$ at $\beta = 2$, $D = 25600$, and by $6\times 10^{-8}$ at $\beta = 8$, $D = 25600$, far below the offsets in Eq.~\eqref{eq:offset_det}. The method reproduces $\mathrm{E}[\bar C_2] = 1/3$ and $\mathrm{E}[\bar C_3] = 241/6-4\pi^2$ at $\beta = 2$. 

The second is the sampled chain of Sec.~\ref{sm:sampled}. A weighted least-squares fit of $\mathrm{E}[\bar C_D]/D - P(\beta) = a/D$ to the ten sizes, with $P(\beta)$ fixed at its exact value and $a$ the only parameter, gives the offsets of Table~\ref{tab:offset}. All eight are within two standard errors of $-1/2$. The same weighted fit with a free intercept, $P_\mathrm{P} + b_\mathrm{P}/D$, applied to independent phases gives the open symbols of Fig.~\ref{fig:A1.0}(c), $P_{\mathrm P}=0.2363(1)$, $0.3232(1)$, $0.3555(1)$, $0.3939(1)$, $0.4242(1)$, $0.4412(1)$, $0.4592(1)$, and $0.4689(1)$ for $\beta=1/2,1,4/3,2,3,4,6,8$, with $\chi^2$ between $3.9$ and $11.4$ on 8 degrees of freedom.

\begin{table}[htb]
\setlength{\tabcolsep}{7pt}
\caption{Offset $a$ fitted to the sampled circular ensemble, $\mathrm E[\bar C_D]/D=P(\beta)+a/D$, with its standard error and the $\chi^2$ of the fit (9 degrees of freedom).}\label{tab:offset}
\begin{tabular}{@{}lcccccccc@{}}
\toprule
$\beta$ & $1/2$ & $1$ & $4/3$ & $2$ & $3$ & $4$ & $6$ & $8$\\
\midrule
$a$ & $-0.478$ & $-0.490$ & $-0.511$ & $-0.507$ & $-0.496$ & $-0.497$ & $-0.503$ & $-0.496$\\
s.e. & $0.012$ & $0.010$ & $0.009$ & $0.007$ & $0.006$ & $0.005$ & $0.003$ & $0.003$\\
$\chi^2$ & $10.1$ & $5.9$ & $7.4$ & $6.6$ & $9.6$ & $9.5$ & $11.7$ & $3.2$\\
\bottomrule
\end{tabular}
\end{table}

\subsection{Estimators and tests for the models}\label{sm:estimators}
Every quantity is first averaged over the initial states of a block, then over the blocks of an operator, yielding a single value $x_o$ per operator. The estimate is $m = n_{\text{op}}^{-1}\sum_o x_o$, with $n_{\text{op}} = 200$, and its standard error is $\sigma = \mathrm{sd}_o(x_o)/\sqrt{n_{\text{op}}}$. The operator is the unit because the initial states of one operator share its spectrum. All error bars of the model data in the Letter are one standard error. For a prediction $x^{\mathrm{th}}$, the residual is $\chi = (m - x^{\mathrm{th}})/\sigma$. The number of operators and the acceptance bands ($\abs{\chi}\leq 3$ for a point and a probability of at least $0.01$ for the sum $\chi^2$ over a family of points) were fixed before the final data set was generated. 

\paragraph{Reference with matched weights.} For every block of every chaotic model, we also draw one circular-ensemble spectrum of the same $D$ and class and give it the weights of the model, transferred by the rank of the eigenphase in $[0,2\pi)$: the level with the $r$-th smallest phase of the model gives its weight to the level with the $r$-th smallest phase of the reference. All quantities are then computed for the model and the reference using the same weights and code. Their difference, formed operator by operator before averaging, is free of the fluctuations of the weights but carries those of the reference spectrum. We write $\Delta$ for this paired difference of the spread divided by $D$ and $d$ for that of the last-site occupation. This reference is the one labeled KN in Fig.~\ref{fig:A4}(a); in Fig.~\ref{fig:A4}(c), KN denotes instead the sampled circular ensemble of Sec.~\ref{sm:sampled}. 

\paragraph{Dependence between sizes.} The circuits use independent gates at every size. The four sizes of a kicked top share their 200 sets of couplings (Sec.~\ref{sm:models}), so their residuals need not be independent, although the initial states are. For the last-site occupation, the correlation coefficient of the operator means between two sizes of the same top lies between $-0.11$ and $+0.18$ (18 pairs of sizes, 200 operators each, for which independent values would give a standard deviation of $0.07$). Including this covariance changes the combined standard errors accross the four sizes of a top by less than $10\%.$

\subsection{Mean spread of the models}\label{sm:plateautests}
Figure~\ref{fig:A3}(b) shows the residuals. Of the 27 residuals with respect to $P(\beta) - 1/(2D)$, 25 are within two standard errors and all within $2.31$; their sum of squares is $29.3$. At the largest size of each model, $\abs{m - P(\beta) + 1/(2D)} \leq 5.4 \times 10^{-5}$, with standard errors between $1.6\times 10^{-5}$ and $3.8 \times 10^{-5}$. Without the offset, i.e., against $P(\beta)$ alone, every point lies between 6 and 133 standard errors below the prediction. The reference with matched weights follows $P(\beta) - 1/(2D)$ at all 27 points (largest residual $2.32$). The paired difference $\Delta$ is at most $2.5$ standard errors at 26 points. The exception is the orthogonal circuit at $D = 512$, with $\Delta = (1.75 \pm 0.54) \times 10^{-4}$; at the neighboring sizes, $\Delta$ is $1.1$ and $-0.3$ standard errors, and the model's own residual with respect to $P(\beta) - 1/(2D)$ is $0.76$. 

The spread of a single initial state fluctuates around these means. Over all initial states and operators of a point, $\mathrm{Var}(\bar C_D)/D$ is $0.043$--$0.045$, $0.023$--$0.024$, and $0.0097$--$0.0099$ for $\beta = 1,2,4$, with no systematic dependence on $D$, so the relative fluctuation of $\bar C_D$ decreases as $D^{-1/2}$. The ratio of this variance to that of the reference with matched weights is within $2.4$ standard errors of one at all 27 points. 

\subsection{Last site of the models}\label{sm:farendtests}
Table~\ref{tab:farend} gives the data of Fig.~\ref{fig:A2}(a): the last-site occupation $y$ with its standard error $\sigma_y$, the exact value $y_{\mathrm{ex}}$ at the same $D$, the residual $\chi_y=(y-y_{\mathrm{ex}})/\sigma_y$, and the paired difference $d$. Where $D$ is not a power of two, $y_{\mathrm{ex}}$ is interpolated in $\ln D$; withholding every second lattice size changes the interpolated values by less than $5\times10^{-4}$. The sum of $\chi_y^2$ is $17.1$ on 10 points for $\beta=1$ and $5.3$ on 10 points for $\beta=2$. For $\beta=4$, it is $47.3$ on 7 points, which is outside the band. The excess is carried by the symplectic kicked top. Its four sizes lie above the exact chain by $(3.1\pm1.8)$, $(4.9\pm1.9)$, $(6.9\pm1.7)$, and $(6.3\pm1.5)\times10^{-3}$ at $D=100,200,400,800$, the last two being the only residuals of the table beyond three standard errors. Their inverse-variance mean is $(5.5\pm0.9)\times10^{-3}$ on $y\approx0.54$, that is, $(1.0\pm0.2)\%$, with no detectable dependence on $D$. The operator means at different sizes are nearly uncorrelated (next paragraph), and a generalized least-squares mean that includes their covariance gives the same value, $(5.5\pm0.9)\times10^{-3}$. The symplectic circuit gives $(-1.6\pm0.8)\times10^{-3}$.

The paired difference is a control, not a more sensitive test: $\sigma_d$ exceeds $\sigma_y$ at all 27 points, by factors between $1.2$ and $1.7$, because the reference contributes the fluctuations of its own spectrum. Its use is that both terms carry the same weight, are computed by the same code, and involve no interpolation. The sum of $\chi_d^2$ is $10.9$ on 10 points for $\beta=1$, $9.2$ on 10 for $\beta=2$, and $22.1$ on 7 for $\beta=4$. For the symplectic top, the inverse-variance mean of $d$ over the four sizes is $(5.1\pm1.2)\times10^{-3}$, consistent with the excess over the exact chain. The excess is therefore a property of the quasienergies of this model at these couplings and sizes, and not of the weights or of the numerical procedure. Its spacing ratio does not show it (Fig.~\ref{fig:rbar}), and neither does its mean spread. Section~\ref{sm:sff} relates it to the form factor of the top at the first few periods. The orthogonal top lies below the exact chain across its four sizes by $(-3.1\pm1.1)\times10^{-3}$ on average; the same estimate yields a deficit of this sign and order.

\begin{table}[htb]
\setlength{\tabcolsep}{7pt}
\caption{Last-site occupation of the models. $y$: mean of $D\bar p_{D-1}$ with standard error $\sigma_y$; $y_{\mathrm{ex}}$: exact chain; $\chi_y=(y-y_{\mathrm{ex}})/\sigma_y$; $d$: paired difference with the reference carrying the same weights, with standard error $\sigma_d$ and $\chi_d=d/\sigma_d$; $p_{\mathrm{prof}}$: sign-flip test of the profile difference in eight bins (Sec.~\ref{sm:profiles}).}\label{tab:farend}
\begin{tabular}{@{}lrcccccccc@{}}
\toprule
model & $D$ & $y$ & $10^3\sigma_y$ & $y_{\mathrm{ex}}$ & $\chi_y$ & $10^3d$ & $10^3\sigma_d$ & $\chi_d$ & $p_{\mathrm{prof}}$\\
\midrule
top, $\beta=1$ & 100 & 0.0645 & 2.9 & 0.0725 & $-2.77$ & $-8.0$ & 3.9 & $-2.07$ & 0.84 \\
 & 200 & 0.0547 & 2.6 & 0.0580 & $-1.29$ & $+1.3$ & 3.4 & $+0.39$ & 0.76 \\
 & 400 & 0.0445 & 2.3 & 0.0470 & $-1.12$ & $-3.8$ & 3.2 & $-1.16$ & 0.47 \\
 & 800 & 0.0367 & 1.5 & 0.0386 & $-1.21$ & $-3.4$ & 2.6 & $-1.27$ & 0.60 \\
\midrule
top, $\beta=2$ & 100 & 0.2859 & 3.0 & 0.2848 & $+0.40$ & $-1.0$ & 4.3 & $-0.24$ & 0.04 \\
 & 200 & 0.2616 & 3.1 & 0.2644 & $-0.92$ & $-3.4$ & 4.4 & $-0.77$ & 0.15 \\
 & 400 & 0.2490 & 2.8 & 0.2477 & $+0.44$ & $+6.8$ & 3.8 & $+1.81$ & 0.61 \\
 & 800 & 0.2321 & 2.6 & 0.2337 & $-0.62$ & $-2.8$ & 3.9 & $-0.70$ & 0.21 \\
\midrule
top, $\beta=4$ & 100 & 0.5612 & 1.8 & 0.5580 & $+1.71$ & $+6.8$ & 2.4 & $+2.79$ & 0.22 \\
 & 200 & 0.5519 & 1.9 & 0.5470 & $+2.54$ & $+4.4$ & 2.7 & $+1.66$ & 0.49 \\
 & 400 & 0.5454 & 1.7 & 0.5386 & $+4.16$ & $+5.4$ & 2.4 & $+2.24$ & 0.51 \\
 & 800 & 0.5382 & 1.5 & 0.5319 & $+4.14$ & $+3.7$ & 2.2 & $+1.66$ & 0.65 \\
\midrule
circuit, $\beta=1$ & 64 & 0.0868 & 4.5 & 0.0845 & $+0.51$ & $-0.2$ & 5.8 & $-0.03$ & 0.37 \\
 & 128 & 0.0614 & 3.1 & 0.0668 & $-1.77$ & $-0.6$ & 4.0 & $-0.15$ & 0.30 \\
 & 256 & 0.0558 & 4.0 & 0.0537 & $+0.53$ & $+4.1$ & 4.8 & $+0.86$ & 0.36 \\
 & 512 & 0.0459 & 3.7 & 0.0438 & $+0.58$ & $+1.9$ & 4.5 & $+0.42$ & 0.0004 \\
 & 1024 & 0.0384 & 2.6 & 0.0360 & $+0.91$ & $+4.8$ & 3.9 & $+1.23$ & 0.86 \\
 & 2048 & 0.0312 & 2.7 & 0.0299 & $+0.47$ & $+3.7$ & 3.7 & $+1.02$ & 0.68 \\
\midrule
circuit, $\beta=2$ & 64 & 0.2969 & 4.5 & 0.3004 & $-0.76$ & $-6.0$ & 6.4 & $-0.95$ & 0.57 \\
 & 128 & 0.2799 & 4.6 & 0.2770 & $+0.63$ & $+0.8$ & 6.3 & $+0.13$ & 0.05 \\
 & 256 & 0.2613 & 4.0 & 0.2581 & $+0.78$ & $+2.8$ & 5.5 & $+0.50$ & 0.47 \\
 & 512 & 0.2443 & 4.5 & 0.2425 & $+0.41$ & $+1.2$ & 6.0 & $+0.20$ & 0.26 \\
 & 1024 & 0.2325 & 4.1 & 0.2292 & $+0.80$ & $+3.0$ & 5.8 & $+0.52$ & 0.44 \\
 & 2048 & 0.2224 & 3.9 & 0.2179 & $+1.15$ & $+9.1$ & 5.0 & $+1.81$ & 0.59 \\
\midrule
circuit, $\beta=4$ & 64 & 0.5656 & 2.1 & 0.5668 & $-0.62$ & $-2.7$ & 2.8 & $-0.96$ & 0.91 \\
 & 256 & 0.5419 & 1.3 & 0.5438 & $-1.36$ & $-3.3$ & 2.1 & $-1.59$ & 0.03 \\
 & 1024 & 0.5284 & 1.3 & 0.5299 & $-1.15$ & $-1.3$ & 2.1 & $-0.60$ & 0.97 \\
\bottomrule
\end{tabular}
\end{table}

\subsection{Short-time form factor and the last site}\label{sm:sff}
By End Matter Eq.~\eqref{em:last}, the last-site occupation depends on the quasienergies only through the weights $b_a\propto\abs{L'(z_a)}^{-2}$. For distinct eigenphases, the Fourier series $\ln\abs{1-e^{i\theta}}=-\sum_{t\geq1}\cos(t\theta)/t$, valid for $\theta\neq0$, gives
\begin{equation}\label{eq:lnb}
  \ln b_a=\mathrm{const}+2\sum_{t\geq1}\frac1t\left\{\mathrm{Re}\!\left[z_a^{-t}\,\mathrm{tr}\,U^t\right]-1\right\},
  \qquad \mathrm{tr}\,U^t\equiv\sum_{a'}z_{a'}^t ,
\end{equation}
with a constant that does not depend on $a$. The traces of the first few powers of $U$ thus enter the last-site occupation with the largest weights, and these traces are not universal. Table~\ref{tab:sff} lists the form factor $K(t)=\mathrm E\abs{\mathrm{tr}\,U^t}^2$ of the models at $t=1,2,3$, for one block and with Kramers pairs counted once. In the circular ensembles, $K(t)\simeq2t/\beta$ for $t\ll D$. The symplectic top has $K(1)\approx1.0$, twice the circular value, and the orthogonal top has $K(1)\approx0.4$, five times below it.

To estimate the effect on the last site, we deform the circular ensemble by the factor $\exp[\sum_{t=1}^{5}\epsilon_tk_t]$, with $k_t=\abs{\mathrm{tr}\,U^t}^2$. Among the ensembles with prescribed form factor at $t\leq5$, this one has the least relative entropy with respect to the circular ensemble. To first order in $\epsilon_t$, the mean of any function $y_z$ of the phases shifts by
\begin{equation}\label{eq:response}
  \delta\,\mathrm E[y_z]=\sum_{t,s=1}^{5}\mathrm{Cov}(y_z,k_t)\,\bigl(\mathsf C^{-1}\bigr)_{ts}\,\delta K(s),\qquad
  \mathsf C_{ts}=\mathrm{Cov}(k_t,k_s),
\end{equation}
where $\delta K$ is the prescribed shift of the form factor. We take for $y_z$ the average of $D\bar p_{D-1}$ over the weights at fixed phases, for $\delta K$ the difference between the form factors of the model and of the circular ensemble, and we sample the covariances in the circular ensemble (between 400 spectra at $D=800$ and 5000 at $D=64$, with 40 weight vectors each). The last two columns of Table~\ref{tab:sff} compare this estimate with the measured deviation $y-y_{\mathrm{ex}}$ of Table~\ref{tab:farend}. The estimate reproduces the sign and magnitude of the deviations for both tops. Over the 12 points of the table, the sum of squared deviations, in units of their standard errors, is $57.2$ without the estimate and $5.4$ with it, including the estimate's errors. The inverse-variance mean of the estimate over the four sizes of the symplectic top is $(5.7\pm0.6)\times10^{-3}$, against the measured $(5.5\pm0.9)\times10^{-3}$; for the orthogonal top, the corresponding numbers are $(-5.3\pm0.6)\times10^{-3}$ and $(-3.1\pm1.1)\times10^{-3}$.

Equation~\eqref{eq:response} is a first-order estimate for a deformation that is not small, since $\delta K(1)/K(1)\approx1$ for the symplectic top; it is truncated at five periods; and the quasienergies of a model need not be a least-entropy deformation of the circular ensemble. It indicates the origin of the deviations and is not a prediction. With this caveat, the last-site excess of the symplectic top and the smaller deficit of the orthogonal top are accounted for by the nonuniversal form factor of the kicked tops at the first few periods, to which the mean spread and the spacing ratio are not sensitive at the present accuracy.

\begin{table}[htb]
\setlength{\tabcolsep}{7pt}
\caption{Form factor $K(t)=\mathrm E\abs{\mathrm{tr}\,U^t}^2$ of the models at $t=1,2,3$ (circular ensembles: $2t/\beta$ for $t\ll D$), first-order estimate $\delta y$ of the shift of the last-site occupation from Eq.~\eqref{eq:response}, and measured deviation $y-y_{\mathrm{ex}}$ from the exact chain. Parentheses and $\pm$ give one standard error; the error of $\delta y$ combines those of the measured form factor and of the sampled covariances.}\label{tab:sff}
\begin{tabular}{@{}lrccccc@{}}
\toprule
model & $D$ & $K(1)$ & $K(2)$ & $K(3)$ & $10^3\,\delta y$ & $10^3(y-y_{\mathrm{ex}})$\\
\midrule
top, $\beta=1$ & 100 & $0.36(2)$ & $3.39(17)$ & $5.69(30)$ & $-6.5\pm0.9$ & $-8.1\pm2.9$ \\
 & 200 & $0.37(2)$ & $3.89(23)$ & $5.61(32)$ & $-4.1\pm1.3$ & $-3.3\pm2.6$ \\
 & 400 & $0.38(2)$ & $4.15(26)$ & $5.28(29)$ & $-5.2\pm1.6$ & $-2.6\pm2.3$ \\
 & 800 & $0.40(3)$ & $4.92(34)$ & $6.54(40)$ & $-1.3\pm2.2$ & $-1.9\pm1.5$ \\
top, $\beta=2$ & 100 & $0.74(3)$ & $3.09(13)$ & $2.63(14)$ & $+0.3\pm0.9$ & $+1.2\pm3.0$ \\
top, $\beta=4$ & 100 & $0.98(6)$ & $1.26(8)$ & $1.61(13)$ & $+6.6\pm1.1$ & $+3.1\pm1.8$ \\
 & 200 & $0.91(6)$ & $1.41(10)$ & $1.76(12)$ & $+4.1\pm0.9$ & $+4.9\pm1.9$ \\
 & 400 & $1.01(8)$ & $1.75(12)$ & $1.93(16)$ & $+6.7\pm1.6$ & $+6.9\pm1.7$ \\
 & 800 & $1.03(8)$ & $1.65(12)$ & $1.86(12)$ & $+7.7\pm1.7$ & $+6.3\pm1.5$ \\
circuit, $\beta=1$ & 64 & $2.35(18)$ & $3.62(25)$ & $5.92(38)$ & $+2.4\pm1.2$ & $+2.3\pm4.5$ \\
circuit, $\beta=2$ & 64 & $0.94(7)$ & $2.37(16)$ & $3.07(22)$ & $+0.6\pm1.2$ & $-3.4\pm4.5$ \\
circuit, $\beta=4$ & 64 & $0.37(3)$ & $0.97(6)$ & $1.55(10)$ & $-2.0\pm0.6$ & $-1.3\pm2.1$ \\
\bottomrule
\end{tabular}
\end{table}

\subsection{Profiles}\label{sm:profiles}
\paragraph{Models against the matched reference, Fig.~\ref{fig:A4}(a).} For each model at its largest size, the difference between the occupation of the model and that of the reference with matched weights, in units of $1/D$, is averaged over the sites of each of eight bins of $n/D$, then over the blocks of an operator, and then over operators. The largest bin difference is $5.9\times10^{-4}$ (unitary top, $D=800$, bin centered at $n/D=0.81$), which is $2.6$ standard errors. A joint test uses all 27 points. With $d_{b,o}$ the difference in bin $b$ for operator $o$, $\bar d_b$ its mean over operators, and $v_b=\sum_od_{b,o}^2/n_{\mathrm{op}}^2$, the statistic is $\sum_b\bar d_b^{\,2}/v_b$. Its null distribution is obtained by flipping the sign of the whole eight-component vector of each operator, which leaves every $v_b$ and the correlations between bins unchanged; $p_{\mathrm{prof}}$ in Table~\ref{tab:farend} is the fraction of $65\,536$ random sign assignments with a statistic at least as large as observed. At 26 points, $p_{\mathrm{prof}}\geq0.03$. The orthogonal circuit at $D=512$ has $p_{\mathrm{prof}}=3.5\times10^{-4}$; this is the point with the largest paired difference $\Delta$ of the mean spread (Sec.~\ref{sm:plateautests}), as it must be, since the mean spread is the first moment of the profile.

\paragraph{Independent phases, Fig.~\ref{fig:A4}(b,c).} The curves are site-by-site means over the $2000$ spectral measures at $D=2048$, and the symbols are means over 32 bins. At the first site, the data give $D\,\mathrm E[\bar p_0]=2.998(2)$, $1.999(1)$, and $1.4989(4)$ for the weights of $\beta=1,2,4$, against $2.9971$, $1.9990$, and $1.4996$ from Eq.~\eqref{em:origin}, which holds for any phases. The difference from the circular-ensemble profile in Fig.~\ref{fig:A4}(c) sums to zero. It is negative for $n/D<0.63$, $0.67$, and $0.68$, with minima $-0.088$, $-0.079$, and $-0.058$ at $n/D=0.27$, $0.34$, and $0.37$, and it reaches $+0.22$, $+0.30$, and $+0.20$ at the last site. Its first moment, $D^{-2}\sum_nn(\cdot)$, is $0.0200$, $0.0177$, and $0.0124$, which equals the gap $P_{\mathrm P}-P(\beta)$ between the dashed and solid lines of Fig.~\ref{fig:A3}(a) up to terms of order $1/D$.

\section{Meaning of $\beta$, observables, and localization in energy space}\label{sm:beta}

\subsection{The Dyson index as a coupling constant}\label{sm:beta-coupling}
The circular $\beta$-ensemble is the Gibbs measure of charges on the unit circle with logarithmic repulsion at inverse temperature $\beta$. The three circular ensembles selected by Wigner--Dyson symmetry have $\beta=1,2,4$. For any $\beta>0$, the phase law is realized by the chain of Killip and Nenciu~\cite{Killip}, in which $c=\beta/2$ enters only through the reflection laws of End Matter Eq.~\eqref{em:kn}. The mean reflection probability at distance $r$ from the far end is $1/(1+cr)$, so that $\eta^2=2/\beta$ is the coefficient of the marginal inverse-distance decay discussed in Sec.~\ref{ssec:palm}. The Dyson index is thus an inverse coupling constant of the scattering chain. At fixed $D>1$, as $\beta\to0$, the reflection probabilities tend to one, the phases approach independent uniform variables, and the Dirichlet weight vector concentrates on a randomly chosen vertex of the simplex. Thus, $\bar p_0\to1$ in probability. Taking $D\to\infty$ first at fixed $c>0$ gives $\mathrm E[\bar p_0]\sim(1+1/c)/D$. As $\beta\to\infty$, the chain becomes transparent, the phases form a rigid lattice, the weights become uniform, and the profile becomes flat, with $P(\beta)\to1/2$ and $y(\infty)\to1$.

For the circular-ensemble model of a chaotic drive with a symmetry-matched initial state, $\beta$ specifies both the exponent of level repulsion and twice the Dirichlet parameter of the weights. The Killip and Nenciu law ties the two together. The return probability in End Matter Eq.~\eqref{em:origin} depends only on $c$. The lower bound of End Matter Sec.~\ref{em:endpoint} uses only the weights at fixed phases, so $y(D)\geq D(c-1)/(Dc-1)$ for $c>1$ holds for any law of distinct phases that is independent of the weights.

\subsection{Observables}\label{sm:observables}
The stationary Krylov occupations and their moments are functionals of the spectral measure $\sum_aw_a\delta_{z_a}$ of the initial state. Its Fourier coefficients are the return amplitudes
\begin{equation}\label{eq:return}
 \mathcal{A}(t)=\langle\psi_0|U^t|\psi_0\rangle=\sum_aw_az_a^t,
\end{equation}
which can be measured interferometrically. For $D$ distinct occupied eigenphases, the exact complex amplitudes $\ \mathcal A(1),\ldots, \mathcal A(D)$ determine the phases and their weights. Use $\mathcal A(0)=1$ and $\mathcal A(-t)=\overline{\mathcal A(t)}$ to form the $(D+1)\times(D+1)$ matrix $T_{mn}= \mathcal A(n-m)$, with $0\le m,n\le D$. For a polynomial $p(z)=\sum_n p_nz^n$, its quadratic form is $\sum_a w_a|p(z_a)|^2$. The one-dimensional null space therefore gives the monic polynomial $\prod_a(z-z_a)$. Its roots yield the eigenphases, and the first $D$ moments then yield the weights via a linear solve. A Szeg\H{o} recursion on the same moments gives the reflection amplitudes. With noisy data, closely spaced phases and weakly excited modes limit the accuracy of the reconstruction.

The time average of $\abs{\mathcal A(t)}^2$ is the return probability $\bar p_0=\sum_aw_a^2$. By End Matter Eq.~\eqref{em:origin}, $D\mathrm E[\bar p_0]\to1+1/c$, which equals $3$, $2$, and $3/2$ for $\beta=1,2,4$. This observable measures the weight statistics and, like the mean spread $P(\beta)D$, varies smoothly with $\beta$.

The transition is carried by the last site. By End Matter Eq.~\eqref{em:last}, $\bar p_{D-1}$ is the harmonic mean of the overlaps with weights $b_a$ fixed by the quasienergies. For a fixed eigenvalue label, each overlap has the law ${\rm Beta}(c,(D-1)c)$, so $\mathrm E[(Dw_a)^{-1}]=(Dc-1)/[D(c-1)]$ for $c>1$, while the mean inverse overlap diverges for $c\leq1$. With $1/\infty=0$, the endpoint theorem for the matched circular ensembles, Eq.~\eqref{eq:end}, can therefore be written as
\begin{equation}\label{eq:harmonic}
 \lim_{D\to\infty}y(D)=\frac{1}{\lim_{D\to\infty}\mathrm E[(Dw_a)^{-1}]}.
\end{equation}
Small overlaps enter this harmonic mean with a large inverse weight. For the matched ensembles, the density of $w_a$ behaves as $w^{c-1}$ near zero. In the orthogonal class, it diverges there, with the Porter--Thomas law for $Dw_a$ in the large-$D$ limit. In the unitary class, the density is finite at zero and the inverse moment diverges logarithmically as a lower weight cutoff is removed. These statements concern the weight distribution at fixed $D>1$. The slow dependence of the critical endpoint on $D$ also involves the eigenphases. In the symplectic class, the weight density vanishes linearly at zero, and the finite inverse moment gives $y(\infty)=1/2$.

The endpoint theorem also controls fluctuations from one drive to the next. Let $ \hat y_D=D\bar p_{D-1}$ for an individual realization, so $y(D)=\mathrm E[\hat y_D]$. Write $\hat y_D^{-1}=\sum_a b_a/(Dw_a)$, with $b_a\geq0$ and $\sum_a b_a=1$. By independence of the phases and the exchangeable weights,
\begin{equation}\label{eq:end-inverse}
 \mathrm E[\hat y_D^{-1}\mid z_1,\ldots,z_D]=\mathrm E[(Dw_a)^{-1}] .
\end{equation}
For $c>1$, set $\tilde{y}_D=(\mathrm E[\hat y_D^{-1}])^{-1}=D(c-1)/(Dc-1)$. Expanding a square and using Eq.~\eqref{eq:end-inverse} gives
\begin{equation}\label{eq:end-gap}
 \mathrm E\!\left[\frac{(\hat y_D-\tilde y_D)^2}{\hat y_D}\right]=y(D)-\tilde y_D .
\end{equation}
The right-hand side tends to zero by Eq.~\eqref{eq:end}. Cauchy--Schwarz now bounds $\mathrm E\abs{\hat y_D-\tilde y_D}$ by $[y(D)(y(D)- \tilde y_D)]^{1/2}$. For $c\leq1$, $\hat y_D\geq 0$ and $\mathrm E[\hat y_D]\to0$. Thus, in both regimes,
\begin{equation}\label{eq:end-sample}
 D\bar p_{D-1}\longrightarrow (1-2/\beta)_+
 \qquad\text{in }L^1\text{ and in probability}.
\end{equation}
The statement applies to the occupation after summing over all eigenmodes of a single drive. An individual eigenmode, sampled with its initial weight, has the fluctuating profile given by Eq.~\eqref{eq:gibbs}.

The convergence in Eq.~\eqref{eq:end-sample} is slow. We sampled the matched ensemble, with one matrix of Eq.~\eqref{eq:CMV} per drive, with the laws of Eq.~\eqref{eq:verb}, a launch at the first site, and 1000 drives per size (400 at $D = 512$). At $\beta = 4$, the standard deviation of $\hat y_D$ over drives is $0.159$, $0.150$, $0.128$, and $0.123$ at $D = 64, 128, 256, 512$, around means $0.575(5)$, $0.552(5)$, $0.549(4)$, and $0.530(6)$ that agree with the exact chain ($0.5668$, $0.5537$, $0.5438$, $0.5360$). The two sides of Eq.~\eqref{eq:end-gap} agree: $0.068(10)$, $0.059(7)$, $0.048(14)$, and $0.041(6)$ against $y(D)-\tilde y_D=0.0629$, $0.0518$, $0.0428$, and $0.0355$. On the exact chain, the bound $[y(D)(y(D)-\tilde y_D)]^{1/2}$ on $\mathrm E|\hat y_D-\tilde y_D|$ is still $0.056$ at $D=2^{20}$, where $y=0.5061$. At $\beta=2$, the standard deviation is $0.214$, $0.180$, $0.156$, and $0.162$ at the same sizes, comparable to the exact means $0.300$, $0.277$, $0.258$, and $0.243$. The convergence also depends on the phase law. For independent uniform phases, End Matter Eq.~\eqref{em:last} still holds and the mean of $\hat y_D$ does not depend on $D$ (Sec.~\ref{sm:numerics}), but its standard deviation over spectra stays between $0.38$ and $0.46$ at $D=256$ and $1024$ for the three weight laws: about two or three closely spaced phases carry most of the weights $b_a$, so the harmonic mean in Eq.~\eqref{em:last} does not average over many levels.

The equality in Eq.~\eqref{eq:harmonic} uses the upper bound from the circular ensemble as well as the harmonic-mean lower bound. For example, at fixed $D$, keep all phases separated except for one pair and send its spacing to zero through positive values. Since $b_a\propto \abs{L'(z_a)}^{-2}$ with $L(z)=\prod_j(z-z_j)$, the pair carries limiting weights $b_1=b_2=1/2$. In this limit, $\hat y_D$ approaches $2Dw_1w_2/(w_1+w_2)$, whose mean is $2c/(2c+1)$. Indeed, $w_1/(w_1+w_2)$ has law $\mathrm{Beta}(c,c)$ and is independent of $w_1+w_2$. This positive answer persists if $D\to\infty$ after the close-pair limit, even when $c\leq1$.

Since the lower bound depends only on the weights, the endpoint can be raised without changing the quasienergy statistics. For example, take $U=U_{\rm orb}\otimes\mathbf 1_2$, with $U_{\rm orb}$ drawn from the orthogonal class and the spin left unresolved. Each orbital quasienergy then occurs twice. A Haar-random complex initial state on the full $2D$-dimensional space has Dirichlet$(2)$ weights on the $D$ distinct quasienergies, and $y(D)\geq D/(2D-1)$ at every $D$.

\subsection{A chain with tunable $\beta$}\label{sm:beta-chain}
The CMV form $U=\mathcal L\mathcal M$ of Sec.~\ref{sm:arnoldi-cmv} is a product of two layers of $2\times2$ blocks $\Theta_k$ acting on neighboring sites. Each block is a lossless beam splitter that keeps the walker at its site with amplitude $\alpha_k^*$ or $-\alpha_k$ and transfers it to the neighboring site with amplitude $\rho_k$. Drawing the amplitudes independently with uniform phases and the laws of End Matter Eq.~\eqref{em:kn}, with a unimodular last coefficient of uniform phase, gives circular-$\beta$ quasienergies and Dirichlet$(\beta/2)$ weights at the first site for every $\beta>0$~\cite{Killip}. Such networks can be built from programmable beam splitters, with single photons or classical light, or in the one-excitation sector of a particle-conserving two-layer qubit circuit~\cite{TruninHuse2025}. For a walker launched at the first site, the time-averaged intensity at site $n$ equals $\bar p_n$, because $\abs{\chi_n}=\abs{\phi_n}$ on the unit circle. The time-averaged intensity at the last output port can be measured directly. After averaging for $T$ periods, two eigenphases separated by $\Delta\theta$ retain an interference factor of magnitude $\abs{\sin(T\Delta\theta/2)}/[T\abs{\sin(\Delta\theta/2)}]$. The Heisenberg time $D$ corresponds to the mean spacing. Smaller splittings with appreciable interference weight can require longer observation. Optical loss and experimental noise must be sufficiently small to resolve the signal over the chosen time window.

For $\beta>2$, the bound $y(D)\geq D(c-1)/(Dc-1)$ holds at every $D$. A positive finite-size signal also occurs on the depleted side, so identifying the transition requires varying the chain length. Near $\beta=2$, the approach to the limit is slow. In End Matter Eq.~\eqref{em:tilt}, the path accumulates the drift $2/\beta-1$ up to the logarithmic distance $s$, which is close to $\ln D$ at the last site. Comparing the accumulated drift with the Brownian fluctuation of order $\sqrt{(4/\beta)s}$ suggests a crossover for $\abs{1-2/\beta}\sqrt{\ln D}=O(1)$ near $\beta=2$. This is a continuum estimate for the rounding on a finite chain.

\subsection{Localization in energy space}\label{sm:energy}
Periodic driving can localize a system in energy. For the kicked rotor, the Floquet eigenvalue problem maps onto a one-dimensional Anderson model on the lattice of angular momenta, with pseudorandom site energies and hopping that is independent of the angular-momentum index~\cite{FishmanGrempelPrange1982}. In its dynamically localized regime, away from quantum resonances, the Floquet states are exponentially localized, and the kinetic energy stays bounded. Adding two incommensurate modulation frequencies to the periodic kicks produces a different problem, which maps onto a three-dimensional Anderson model and has been observed with cold atoms~\cite{Chabe2008}. The Krylov chain differs in that its reflection probability decays as the inverse distance.

This decay arises in a ring threaded by a magnetic flux that grows linearly in time~\cite{GefenThouless1987,BlatterBrowne1988}. Take a single-channel ring of circumference $L=2\pi R$ and electrons of mass $m$. Without scatterers, the levels are $E_j(\varphi)=E_0(j+\varphi)^2$, with $E_0=\hbar^2/(2mR^2)$, integer $j$, and $\varphi=\Phi/\Phi_0$. Take $\dot\varphi>0$. One flux quantum restores the Hamiltonian up to a large gauge transformation. With $S \ket{j} =\ket{j-1}$, one has $H(\varphi+1)=S H(\varphi)S^\dagger$. If $\mathcal U(T_\Phi,0)$ is the physical propagator over $T_\Phi=1/\dot\varphi$, the one-period map in the initial gauge is $U_F=S^\dagger\mathcal U(T_\Phi,0)$. The authors of Ref.~\cite{GoldhirschLubinGefen1991} make the same gauge identification for a tight-binding ring in their Eq.~(5). Scatterers open gaps between $j$ and $-j-1$ at $\varphi=1/2$, and between $j$ and $-j$ at $\varphi=0$. Label the adiabatic levels by energy, $k=0,1,2,\ldots$. To leading order in the scattering potential, the gap between levels $k$ and $k+1$ is $2\abs{V_{k+1}}$, and crossings with even and odd $k$ alternate in time. When each crossing is traversed before the next one and off-resonant levels can be neglected, two alternating layers of $2\times2$ Zener matrices describe the evolution in the ordered ladder, with the ground level coupled only to the first excited level~\cite{BlatterBrowne1988,BourgetHowlandJoye2003}. The propagation phases between crossings supplement the two layers. In the quantum-walk representation, they enter the phases of the effective reflection amplitudes~\cite{CGMV2010}. The reflection probability $R_k$ of link $k$ is the probability of following the adiabatic level through the crossing, that is, of Bragg reflection between the two diabatic states.

At crossing $k$, the diabatic level difference changes at the rate $2(k+1)E_0\dot\varphi$, and the Zener formula gives
\begin{equation}\label{eq:zener}
 R_k=1-\exp\left[-\frac{\pi\abs{V_{k+1}}^2}{(k+1)\hbar E_0\dot\varphi}\right].
\end{equation}
For independently and uniformly positioned point scatterers of fixed strengths, the mean of $\abs{V_q}^2$ does not depend on $q>0$. A single delta scatterer of strength $u$ at the origin gives $ V_q = u/L$. For real Gaussian white noise with covariance $\gamma\delta_L(x-x')$ on the ring, the positive Fourier modes $V_q=L^{-1}\int_0^L V(x)e^{-2\pi iqx/L}\,\dd x$ are independent circular complex Gaussians. Thus, $|V_q|^2$ are independent exponentials of mean $\gamma/L$ for $q>0$, while $V_{-q}=V_q^*$. At large $k$, therefore,
\begin{equation}\label{eq:eta-ring}
 \mathrm E[R_k]\simeq\frac{\eta^2}{k+1},\qquad
 \eta^2=\frac{\pi\mathrm E\abs{V_q}^2}{\hbar E_0\dot\varphi}.
\end{equation}
For Gaussian white noise, an electromotive force of magnitude $\abs{\mathcal E}$ gives $\dot\varphi=e\abs{\mathcal E}/h$, and the backscattering length at energy $E$ is $\ell_{\rm b}(E)=2\hbar^2E/(m\gamma)$. Then, $\eta^2=m\gamma L/(\hbar^2 e\abs{\mathcal E})=2(L/\ell_{\rm b})[E/(e\abs{\mathcal E})]$. The work available from the field over one revolution has magnitude $\abs{\mathcal E}$, so $\eta^2$ is twice the number of backscattering lengths traveled, while the field supplies an energy equal to the kinetic energy. This number does not depend on energy, because $\ell_{\rm b}\propto E$. To leading order at large $k$, Eq.~\eqref{eq:eta-ring} agrees with the classical backscattering probability over the distance $v/(2\dot\varphi)$ traveled between successive crossings. That probability decays as $1/k$ because $\ell_{\rm b}$ grows as $v^2$ while the distance grows as $v$.

For Gaussian white noise, Eq.~\eqref{eq:zener} gives the entire reflection law in the isolated-crossing model. Write $q=k+1$ and $Y_q=L\abs{V_q}^2/\gamma$. Since $Y_q$ is exponential of mean one,
$$
 1-R_{q-1}=\exp[-\eta^2 Y_q/q],\qquad \Pr(R_{q-1}>r)=(1-r)^{q/\eta^2},\quad 0\leq r\leq1 .
$$
Thus,
\begin{equation}\label{eq:ring-beta-law}
 R_{q-1}\sim\mathrm{Beta}(1,q/\eta^2),\qquad \mathrm E[R_{q-1}]=\frac{\eta^2}{q+\eta^2}.
\end{equation}
With $c=1/\eta^2$, these are exactly the independent reflection probabilities in Eq.~\eqref{em:kn}, in reverse order. A single delta scatterer gives the same leading inverse-distance decay with a deterministic modulus.

The full half-line ensemble also requires independent uniform reflection angles conditional on all the moduli. For Gaussian disorder, the bare gap angles $\arg V_q$ already have this property. Suppose that, in a fixed phase convention, the effective reflection angle at crossing $q$ is $\pm\arg V_q+\delta_q$, with the sign fixed and the phase shift $\delta_q$ determined by the clean flight phases and the gap moduli. Conditioning on all the moduli then leaves the angles independent and uniform. For the microscopic ring, one must also control propagation between crossings, virtual transitions through other levels, and their possible correlations with the gap angles. For a single point scatterer, the phases of $V_q$ are locked to its position. Ao found delocalization at resonant driving even for a periodic delta comb and analyzed the propagating front for smoother periodic potentials~\cite{Ao1990}.

With the circular phase law and Eq.~\eqref{eq:ring-beta-law}, the spectral measure of the first site on the infinite ladder is singular continuous for $0<\eta^2\leq1$ and pure point for $\eta^2>1$~\cite{Simon_2005_2,brow3,Chhaibi_2025}. Its exact Hausdorff dimension on the continuous side is $1-\eta^2$. At $\eta^2=1$, its dimension is zero, and it has no atoms. By the unitary RAGE theorem, a state with continuous spectral measure leaves every finite set of levels on average in time. For unbounded, increasing-level energies, the time-averaged energy diverges. With pure-point spectral measure, the evolved state has uniformly small probability above level $K$ as $K\to\infty$. Bounding its mean energy requires control of those tails after multiplication by the level energies.

The earliest arrival at a distant level yields a test that requires no assumptions about the phases. Launch the walker at ground level and consider the first time it reaches level $K$. Every one of the first $K$ links must transmit, so only one path contributes to this amplitude. Its probability is
\begin{align}
 p_{\rm front}(K)&=\prod_{q=1}^K(1-R_{q-1})
  =\exp\!\left[-\eta^2\sum_{q=1}^K\frac{Y_q}{q}\right],
 \label{eq:ring-front}\\
 \mathrm E[p_{\rm front}(K)]&=
 \frac{\Gamma(1+\eta^2)\Gamma(K+1)}
   {\Gamma(K+1+\eta^2)}
 \sim\Gamma(1+\eta^2)K^{-\eta^2}. \nonumber
\end{align}
Since $\sum_q {\rm Var}(Y_q)/q^2<\infty$, the centered sum $\sum_q(Y_q-1)/q$ converges almost surely. Hence, $p_{\rm front}(K)\sim C K^{-\eta^2}$ for almost every sample, with a positive sample-dependent $C$. At a fixed stroboscopic phase, the energy $\epsilon_K$ of the $K$-th level above the ground state grows as $E_0 K^2/4$. Along the sequence of front levels reached at full periods, $K$ grows linearly with the number of periods. Their contribution alone therefore obeys
\begin{equation}\label{eq:ring-energy}
\expval{H-E_{\rm gs}} \geq\ \epsilon_K p_{\rm front}(K)
 \ \sim\ (E_0 C/4)K^{2-\eta^2}.
\end{equation}
In particular, for $1<\eta^2<2$, the ideal random-phase ladder has a pure-point spectral measure for the launch state and an unbounded mean energy. For $\eta^2\geq2$, the front bound alone leaves the growth of that moment open. For a smooth potential, the same path gives a finite asymptotic front probability whenever $\sum_{q\ge1}\abs{V_q}^2/q<\infty$, directly from Eq.~\eqref{eq:zener}. The threshold $\eta^2=1$ specifies spectral type and escape of probability in the time average. A dc absorption rate requires a separate calculation of work per unit time. In the presence of a phase-breaking time, the absorption rate and the resulting resistance were estimated in Ref.~\cite{Gefen01121987}.

The authors of Ref.~\cite{GefenThouless1987} constructed the single-channel transfer matrix and obtained energy localization for scattering parameters evaluated near the Fermi level. In footnote~17, they explicitly pointed out that the increase of Zener transmission at high energies can allow an almost localized packet to leak. Equation~\eqref{eq:ring-front} quantifies this possibility for the marginal ladder. Blatter and Browne found exponential localization at constant Zener amplitudes and weaker localization when the amplitudes varied with energy~\cite{BlatterBrowne1988}. A finite energy-space localization length in a bounded-band tight-binding ring was found in Ref.~\cite{LubinGefenGoldhirsch1990}, where the breakdown of isolated two-level crossings at fast sweep was identified. The authors of Ref.~~\cite{GoldhirschLubinGefen1991} subsequently studied the normalized participation ratio of Floquet eigenstates over adiabatic levels. It varied monotonically with disorder and driving rate in a single channel and reached a maximum at intermediate parameters in a multichannel cylinder, where different transverse ladders mix. This observable counts the levels occupied by a single mode. The finite-chain observable below measures intensity shared by opposite ends. In a separate infinite-wire problem, independently random strengths of point scatterers in a static field yield a transition from power-localized states at weak field to extended states at strong field~\cite{DelyonSimonSouillard1984,FrankLarson2022}.

\subsection{The endpoint as a finite-ladder transition}\label{sm:ladder}
Time-averaged transition probabilities between two sites are symmetric. For distinct eigenphases, the time average from site $i$ to site $j$ is $\sum_a \abs{\bra{i}\ket{z_a}}^2\abs{\bra{j}\ket{z_a}}^2$. With degeneracies, the corresponding sum uses $\abs{\bra{j}P_a\ket{i}}^2$ over the distinct spectral projectors $P_a$. In either case, an endpoint measurement can be made in both directions.

The reversal can be shown for each finite realization. Let $G$ be the Floquet matrix in the forward Krylov basis [Eq.~\eqref{eq:yy16}] and let $J$ reverse the order of its basis vectors. The matrix $JG^{\mathsf T}J$ has positive subdiagonal entries in reverse order. Its first-site return amplitudes equal the last-site return amplitudes of $G$ at every time. Substitution in Eq.~\eqref{eq:yy16} gives its reflection amplitudes
\begin{equation}\label{eq:finite-ladder-reverse}
 \alpha'_j=-\alpha_{D-1}\alpha^*_{D-2-j},\quad 0\leq j\leq D-2,
 \qquad \alpha'_{D-1}=\alpha_{D-1}.
\end{equation}
Conditional on the closing phase $\alpha_{D-1}$, the reversed bulk angles remain independent and uniform. Their reflection probabilities have laws $\mathrm{Beta}(1,c(j+1))$. The eigenphases are unchanged, while the launch weights become the last-site spectral weights of the forward chain. The two stationary endpoint intensities therefore coincide realization by realization under this reversal. 

For a $D$-level ladder with these laws and an independent uniform closing phase, Eq.~\eqref{eq:end} gives $\mathrm E[\bar p_{D-1}]=[(1-\eta^2)_++o(1)]/D$ for a launch at the strongly reflecting end. Equation~\eqref{eq:end-sample} gives the corresponding statement for a typical sample. For $\eta^2<1$, the coefficient $1-\eta^2$ equals the dimension of the infinite-ladder spectral measure. At $\eta^2=1$, both numbers vanish while the infinite-ladder measure remains continuous. The finite-chain measurement takes the infinite-time average before increasing $D$. The front probability in Eq.~\eqref{eq:ring-front} instead measures the first arrival as the observation time grows. A direct derivation of the endpoint coefficient from the spectral dimension remains open (Sec.~\ref{ssec:palm}).

\end{widetext}

\end{document}